\documentclass[showkeys, groupaddress, superscriptaddress, preprint]{revtex4-2}
\usepackage[T1]{fontenc}
\usepackage[utf8]{inputenc}%
\usepackage{amsmath}
\usepackage{amssymb}
\usepackage{amsthm}
\usepackage{textcomp}
\usepackage{graphicx}
\usepackage[colorlinks=true, citecolor=blue, urlcolor=blue, linkcolor=black]{hyperref}
\usepackage{textgreek}
\usepackage{verbatim}
\usepackage{placeins}
\usepackage{tabularx}

\begin{abstract}

Magnetic skyrmions are topological spin textures that hold great promise as nanoscale information carriers in non-volatile memory and logic devices. While room-temperature magnetic skyrmions and their current-induced manipulation were recently demonstrated,  the stray field resulting from their finite magnetization as well as their topological charge limit their  minimum  size and  reliable motion in tracks. Antiferromagnetic (AF) skyrmions  allow these limitations to be lifted owing to their vanishing magnetization and net zero topological charge, promising room-temperature, ultrasmall skyrmions, fast dynamics, and insensitivity to external magnetic fields.  While room-temperature AF spin textures have been recently demonstrated, the observation and controlled nucleation of AF skyrmions  operable at room temperature in industry-compatible synthetic antiferromagnetic (SAF) material systems is still lacking. Here we demonstrate that isolated  skyrmions can be stabilized at zero field and room temperature in a fully compensated  SAF. Using X-ray microscopy techniques, we are able to observe  the  skyrmions in the different SAF layers and demonstrate their  antiparallel alignment. The results are  substantiated by micromagnetic simulations and analytical models using experimental parameters, which confirm the chiral SAF skyrmion spin texture and allow the identification of the physical mechanisms that control the SAF skyrmion size and stability. We also demonstrate  the local nucleation of SAF skyrmions via  local current injection as well as ultrafast laser excitations at zero field. These results will enable the utilization of SAF skyrmions in skyrmion-based devices.

\end{abstract}

\begin{document}

\title{Skyrmions in synthetic antiferromagnets and their   nucleation via electrical current and ultrafast laser illumination}

\author{Rom\'eo Juge}
\affiliation{Universit\'e Grenoble Alpes, CEA, CNRS, Grenoble INP, IRIG-SPINTEC, 38054 Grenoble, France}

\author{Naveen Sisodia}
\affiliation{Universit\'e Grenoble Alpes, CEA, CNRS, Grenoble INP, IRIG-SPINTEC, 38054 Grenoble, France}

\author{Joseba Urrestarazu Larra\~naga}
\affiliation{Universit\'e Grenoble Alpes, CEA, CNRS, Grenoble INP, IRIG-SPINTEC, 38054 Grenoble, France}

\author{Qiang Zhang}
\affiliation{Universit\'e Grenoble Alpes, CEA, CNRS, Grenoble INP, IRIG-SPINTEC, 38054 Grenoble, France}

\author{Van Tuong Pham}
\affiliation{Universit\'e Grenoble Alpes, CEA, CNRS, Grenoble INP, IRIG-SPINTEC, 38054 Grenoble, France}

\author{Kumari Gaurav Rana}
\affiliation{Universit\'e Grenoble Alpes, CEA, CNRS, Grenoble INP, IRIG-SPINTEC, 38054 Grenoble, France}

\author{Brice Sarpi}
\affiliation{Synchrotron SOLEIL, L'Orme des Merisiers, 91190 Saint-Aubin, France}

\author{Nicolas Mille}
\affiliation{Synchrotron SOLEIL, L'Orme des Merisiers, 91190 Saint-Aubin, France}

\author{Stefan Stanescu}
\affiliation{Synchrotron SOLEIL, L'Orme des Merisiers, 91190 Saint-Aubin, France}

\author{Rachid Belkhou}
\affiliation{Synchrotron SOLEIL, L'Orme des Merisiers, 91190 Saint-Aubin, France}

\author{Mohamad-Assaad Mawass}
\affiliation{Helmholtz-Zentrum Berlin f\"ur Materialien und Energie, Albert-Einstein-Straße 15, 12489 Berlin, Germany}

\author{Nina Novakovic-Marinkovic}
\affiliation{Helmholtz-Zentrum Berlin f\"ur Materialien und Energie, Albert-Einstein-Straße 15, 12489 Berlin, Germany}

\author{Florian Kronast}
\affiliation{Helmholtz-Zentrum Berlin f\"ur Materialien und Energie, Albert-Einstein-Straße 15, 12489 Berlin, Germany}

\author{Markus Weigand}
\affiliation{Helmholtz-Zentrum Berlin f\"ur Materialien und Energie, Albert-Einstein-Straße 15, 12489 Berlin, Germany}

\author{Joachim Gr\"afe}
\affiliation{Max Planck Institute for Intelligent Systems, Heisenbergstraße 3, 70569 Stuttgart, Germany}

\author{Sebastian Wintz}
\affiliation{Max Planck Institute for Intelligent Systems, Heisenbergstraße 3, 70569 Stuttgart, Germany}

\author{Simone Finizio}
\affiliation{Swiss Light Source, Paul Scherrer Institut, 5232 Villigen, Switzerland}

\author{J\"org Raabe}
\affiliation{Swiss Light Source, Paul Scherrer Institut, 5232 Villigen, Switzerland}

\author{Lucia Aballe}
\affiliation{ALBA Synchrotron Light Facility, 08290 Cerdanyola del Vall\`es, Barcelona, Spain}

\author{Michael Foerster}
\affiliation{ALBA Synchrotron Light Facility, 08290 Cerdanyola del Vall\`es, Barcelona, Spain}

\author{Mohamed Belmeguenai}
\affiliation{Laboratoire des Sciences des Proced\'es et des Mat\'eriaux, CNRS, Universit\'e Paris 13, 93430 Villetaneuse, France}

\author{Liliana Buda-Prejbeanu}
\affiliation{Universit\'e Grenoble Alpes, CEA, CNRS, Grenoble INP, IRIG-SPINTEC, 38054 Grenoble, France}

\author{Justin M. Shaw}
\affiliation{Quantum Electromagnetics Division, National Institute of Standards and Technology, Boulder, CO 80309, USA}

\author{Hans T. Nembach}
\affiliation{Quantum Electromagnetics Division, National Institute of Standards and Technology, Boulder, CO 80309, USA}
\affiliation{Department of Physics, University of Colorado, Boulder, CO 80309, USA}

\author{Laurent Ranno}
\affiliation{Université Grenoble Alpes, CNRS, Institut N\'eel, 38042 Grenoble, France}

\author{Gilles Gaudin}
\affiliation{Universit\'e Grenoble Alpes, CEA, CNRS, Grenoble INP, IRIG-SPINTEC, 38054 Grenoble, France}

\author{Olivier Boulle}
\email{e-mail: olivier.boulle@cea.fr}
\affiliation{Universit\'e Grenoble Alpes, CEA, CNRS, Grenoble INP, IRIG-SPINTEC, 38054 Grenoble, France}

\maketitle

Magnetic skyrmions have raised considerable interest in the last years motivated by the wealth of physical phenomena they evidence at the frontier between topology and magnetism, as well as promising applications in information technology devices~\cite{fertSkyrmionsTrack2013,fertMagneticSkyrmionsAdvances2017,nagaosaTopologicalPropertiesDynamics2013,gobel_beyond_2021,tokura_magnetic_2020,li_magnetic_2021}. Magnetic skyrmions are bi-dimensional local whirling of magnetization with a non-trivial topology. They have a  topological charge $|Q|=1$, \textit{i.e.}, the skyrmion spin texture wraps once around the unit sphere in spin space.  Their small lateral dimensions, down to a few nanometers, topological invariance and fast manipulation by electrical current can be exploited to store and maneuver the information at the nanoscale in memory and logic devices. The recent demonstration of room-temperature skyrmions in technology-relevant ultra-thin ferromagnetic (FM) films lacking inversion symmetry as well as their fast current-induced manipulation  were first important steps toward the practical realization of such devices~\cite{jiangBlowingMagneticSkyrmion2015,wooCurrentdrivenDynamicsInhibition2018,boulleRoomtemperatureChiralMagnetic2016,moreau-luchaireAdditiveInterfacialChiral2016,jugeCurrentDrivenSkyrmionDynamics2019,jiangDirectObservationSkyrmion2017,litziusSkyrmionHallEffect2017}. These multilayer stacks composed of ultra-thin heavy metal{\slash}ferromagnet films combine the necessary ingredients for the  stabilization of  skyrmions, namely perpendicular magnetic anisotropy (PMA), Dzyaloshinskii-Moriya interaction (DMI) and dipolar interactions. However, FM skyrmions suffer from several drawbacks limiting  their implementation in functional devices.
First, the non-local stray fields in FMs  help stabilize skyrmions  but also limit their minimal size to several tens of nanometers at room temperature \cite{buttnerTheoryIsolatedMagnetic2018}. Furthermore, an external magnetic field is very often needed for their stabilization~\cite{buttnerTheoryIsolatedMagnetic2018,bernand-mantelSkyrmionbubbleTransitionFerromagnetic2018,moreau-luchaireAdditiveInterfacialChiral2016,wooObservationRoomtemperatureMagnetic2016}.
Second, their dynamics in tracks is impaired by the skyrmion Hall effect, a direct consequence of their non-zero topological charge~\cite{jiangDirectObservationSkyrmion2017,litziusSkyrmionHallEffect2017}, which deflects the skyrmions from their straight trajectory along the current, pushing them towards the edge of the device where they can be annihilated, resulting in the loss of information.  Third, their non-zero topological charge further hinders the skyrmion motion by creating topological damping \cite{buttnerTheoryIsolatedMagnetic2018,hrabecCurrentinducedSkyrmionGeneration2017}, which strongly limits their velocity to 100 m/s in FMs~\cite{wooCurrentdrivenDynamicsInhibition2018,jugeCurrentDrivenSkyrmionDynamics2019}.   

These problems can be overcome by considering  AF skyrmions, which consist of skyrmions with antiparallel neighboring spins, and therefore opposite topological charges, such as skyrmions in different sublattices  coupled antiferromagnetically.  AF skyrmions with tens of nm in diameter and zero-field room-temperature stability are predicted due to their vanishing magnetization~\cite{jaeschke-ubiergo_stability_2019,buttnerTheoryIsolatedMagnetic2018,bernand-mantelSkyrmionbubbleTransitionFerromagnetic2018,barker_static_2016,jaeschke-ubiergo_stability_2019}. Furthermore, their zero net topological charge cancels the skyrmion Hall effect and simulations predict a straight  trajectory along the current  with  velocities above 1000 m/s~\cite{jin_dynamics_2016,barker_static_2016,tomasello_performance_2017,zhang_antiferromagnetic_2016,shen_current-induced_2019}. Finally, their zero net magnetization makes them insensitive to external magnetic fields.

To implement AF skyrmions in devices, a first  requirement is to demonstrate their room-temperature zero-field stabilization  as well as their controlled nucleation, a challenging task given their vanishing magnetic moment. Progress has been made recently with the demonstration of topological spin textures in antiferromagnets, such as  fractional AF skyrmion lattices   in bulk MnSc$_2$S$_4$ compound at cryogenic temperature~\cite{gao_fractional_2020} and  AF half-skyrmions and bimerons in $\alpha$-Fe$_2$O$_3$ at room temperature \cite{jani_antiferromagnetic_2021}. However, these AF textures were detected in  bulk crystals or epitaxial  films which are complex to grow and hardly compatible with large-scale CMOS integration. In addition, the controlled nucleation of isolated  AF skyrmions at room temperature using local excitation is still lacking.

A more promising approach lies in synthetic antiferromagnets (SAFs). SAFs are composed of two (or an even number) of ultra-thin FM layers  antiferromagnetically coupled through a non-magnetic spacer, via Ruderman-Kittel-Kasuya-Yosida (RKKY)-type interlayer exchange coupling~\cite{parkin_spin_1991,fert_nobel_2008}.  SAF skyrmions thus consist of a pair (or pairs) of antiferromagnetically coupled skyrmions, each in its respective FM layer.  In contrast to bulk AFs, the magnetic properties of ultra-thin films forming a SAF, such as PMA, DMI and interlayer coupling, can be tuned by adjusting the film thickness and  the nature of the material composing the layers, such that the size and stability of the skyrmions can be adjusted. SAF skyrmions are also insensitive to external magnetic fields as long as these do not outweigh the interlayer exchange coupling. Furthermore, SAFs are composed of technology-relevant sputtered films compatible with CMOS integration  as well as standard spintronics devices such as magnetic tunnel junctions for readout.

However,  while  skyrmions in FM can be easily nucleated from a stripe domain state by an external magnetic field, the vanishing magnetic moments of SAF skyrmions makes their nucleation and visualization  challenging. To circumvent this problem, strategies have been considered  where  SAF skyrmions were nucleated  from a multidomain state using either an external magnetic field in a non-compensated SAF~\cite{dohi_formation_2019} or an internal bias field in a SAF exchanged-coupled to a FM bias layer~\cite{legrandRoomtemperatureStabilizationAntiferromagnetic2019}. However, these approaches have important limitations regarding current-induced skyrmion manipulation: the skyrmion Hall effect is mitigated but not suppressed in non-compensated SAF, and the use of a bias FM layer would lead to current shunting and a sensitivity to external magnetic fields. 

Here, we demonstrate that isolated  skyrmions can be stabilized in a pure, compensated SAF at room temperature and zero external field. Using X-ray magnetic microscopy, we are able to visualize the skyrmions in the different  layers constituting the SAF and to confirm their antiparallel alignment. By combining X-ray microscopy, micromagnetic simulations and analytical models, we confirm their internal chiral spin texture and show that their size and stability can be tailored by tuning the thickness of the constituent layers. Local and controlled nucleation of SAF skyrmions is also demonstrated using pulsed current injection as well as ultrafast laser excitation at zero field. These result will enable the use of SAF skyrmions in future logic and memory devices. 

\begin{figure*}[ht!]
\includegraphics[width=0.95\textwidth]{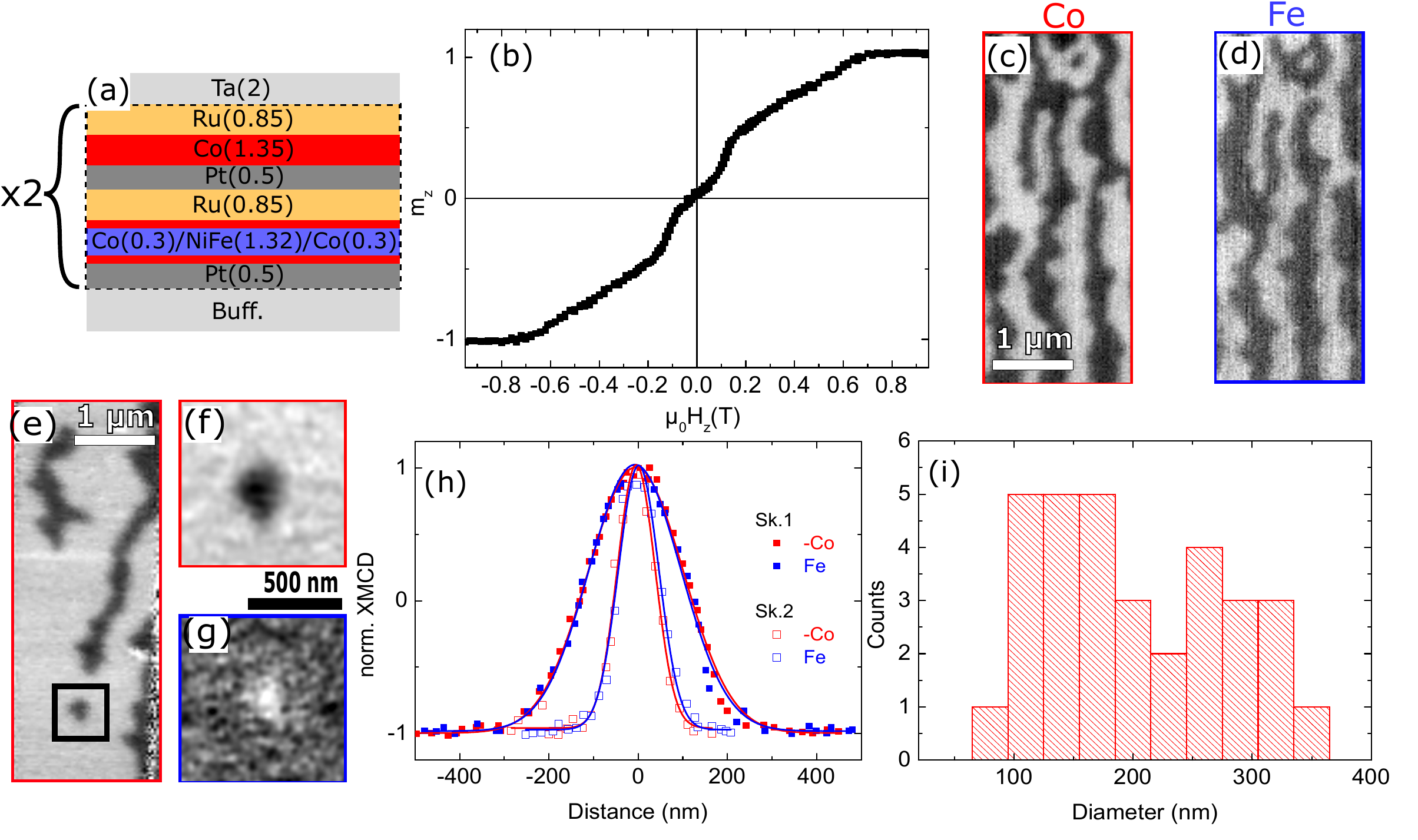}
\centering
\caption{\textbf{Observation of SAF skyrmions $\mid$} (a) Material stack for the SAF. Buff. denotes Ta(3){\slash}Pt(2.5) (thickness in nanometers). (b) Out-of-plane hysteresis loop measured by vibrating sample magnetometry. (c,d) XMCD-STXM images acquired at (c) the Co $L_3$ and (d) the Fe $L_3$ absorption edge after out-of-plane demagnetization. An isolated skyrmion is observed at the top of the image (Sk.2).
(e) XMCD-STXM image acquired at the Co edge  showing an isolated skyrmion (Sk.1). The image was recorded at zero field after the successive application of an external magnetic field  (180 mT)  out of and in the plane of the layers. (f,g) XCMD microcopy image obtained from amplitudes of  ptychography reconstructions  of the skyrmion in the black box in (e) acquired at the Co and Fe edge, respectively. (h) Normalized XMCD signal obtained from line-scans along the skyrmion diameters in (c,d) and (e,g). The signal for Co is inverted and  the solid lines are Gaussian fits. (i) Distribution of skyrmion diameters extracted from different STXM images.}
\label{FIG_SAF_sk_N=2}
\end{figure*}

\subsection*{Observation of  skyrmions in  SAF at room temperature}

The stabilization of SAF skyrmions  requires fine tuning of the magnetic properties of the different FM constituent layers. The latter must combine: (i) PMA; (ii) a large DMI so as to promote spin textures with the same chirality; (iii) a large AF coupling between the FM layers; (iv) equal total magnetic moment to obtain a fully compensated SAF; (v) spin-orbit torques of the same sign in order for both AF-coupled skyrmions to be driven in the same direction.
  
To combine these features,  we optimized a SAF stack with the following composition (see Methods, SAF1): [Pt(0.5){\slash}FM1{\slash}Ru(0.85){\slash}Pt(0.5){\slash}FM2{\slash}Ru(0.85)]$_{\times2}$ where FM1 = Co(0.3){\slash}Ni$_{80}$Fe$_{20}$(1.45){\slash}Co(0.3) and FM2 = Co(1.35) (thickness in nanometers) are antiferromagnetically coupled through the Ru spacer via interlayer exchange coupling (see Fig. \ref{FIG_SAF_sk_N=2}(a)). The thickness of the different ferromagnets composing FM1 and FM2 is adjusted so as to reach magnetic moment compensation and, at the same time, to be close to the spin reorientation (out-of-plane to in-plane) transition (see Fig.~\ref{FIG_SAF_sk_N=2}(b)), which facilitates the skyrmion nucleation by notably reducing the domain wall energy (see Supplementary Information, S1.1.1). Here, we chose to use different FM materials in the two constituent layers. This allows us to distinguish the skyrmions in the Pt{\slash}Co{\slash}NiFe{\slash}Co and the Pt{\slash}Co AF-coupled layers using transmission X-ray microscopy, by adjusting the X-ray energy to the Fe or the Co absorption edge, respectively. From Brillouin light scattering spin wave spectroscopy, we estimate the DMI in the two ferromagnetic layers, FM1 and FM2, to be $D=0.67\pm0.04$ mJ/m$^2$ and $D=0.57\pm0.11$ mJ/m$^2$, respectively (see Supplementary Information, S1.1.2).  The Co{\slash}Ru interfaces enhance the RKKY interaction  and are also expected to add  up to the DMI coming  from the Pt{\slash}Co interfaces \cite{khadka_dzyaloshinskiimoriya_2018,samardak_enhanced_2018}. 

Fig. \ref{FIG_SAF_sk_N=2}(c) and \ref{FIG_SAF_sk_N=2}(d) display X-ray magnetic circular dichroism scanning transmission X-ray microscopy (XMCD-STXM) images of a 2-{\textmu}m-wide SAF track. The images were acquired after demagnetization at the Co and the Fe $L_3$ absorption edge, respectively. The stripe domain structure is explained by the low PMA of the layers. As expected, the magnetic contrast is opposite at the Co and the Fe   edges, confirming that the magnetization points in opposite directions in FM1 and FM2 and that the domains are AF coupled.  A skyrmion can also be seen at the top of the image.  Isolated SAF skyrmions can also be nucleated using large external magnetic field or the  injection of current pulses in the track (see Supplementary Information, S1.2.3). Fig.~\ref{FIG_SAF_sk_N=2}(e) shows a zero field STXM image of a track  with large domains and an isolated skyrmion prepared  after applying successively a magnetic field of 180 mT out of then in the plane of the film.  Fig. \ref{FIG_SAF_sk_N=2}(f) and \ref{FIG_SAF_sk_N=2}(g) show  higher spatial resolution images  of   the SAF skyrmion ,obtained using ptychography reconstruction~\cite{thibault_high-resolution_2008,chapman_coherent_2010,chapman_coherent_2010,shapiro_chemical_2014,donnelly_high-resolution_2016,shi_soft_2016,desjardins_backside-illuminated_2020} and acquired at the Co and the Fe $L_3$ edge, respectively. The opposite magnetic contrast at the two edges demonstrates the antiparallel alignment of the skyrmion magnetizations in FM1 and FM2.  To emphasize the AF coupling,  the XMCD contrasts at the Co and Fe edges  along the diameters of the skyrmions in Fig. \ref{FIG_SAF_sk_N=2}(c,d) and \ref{FIG_SAF_sk_N=2}(f,g) are plotted in Fig. \ref{FIG_SAF_sk_N=2}(h).  For comparison purpose, the Co signal is inverted and both signals are normalized.  Fig. \ref{FIG_SAF_sk_N=2}(h) reveals that both signals superimpose within the resolution limit of the instrument, confirming that this magnetic texture is a SAF skyrmion.  The full width at half maximum of a Gaussian fit provides a measure of the skyrmion diameter at $m_z=0$: 107 nm and 239 nm, respectively. Fig. \ref{FIG_SAF_sk_N=2}(i) shows the distribution of skyrmion diameters obtained after repeating these nucleation procedures using either current or magnetic field at different locations. The skyrmion diameter is on average  204 nm (standard deviation of 85 nm).

\begin{figure*}[t!]
\includegraphics[width=\textwidth]{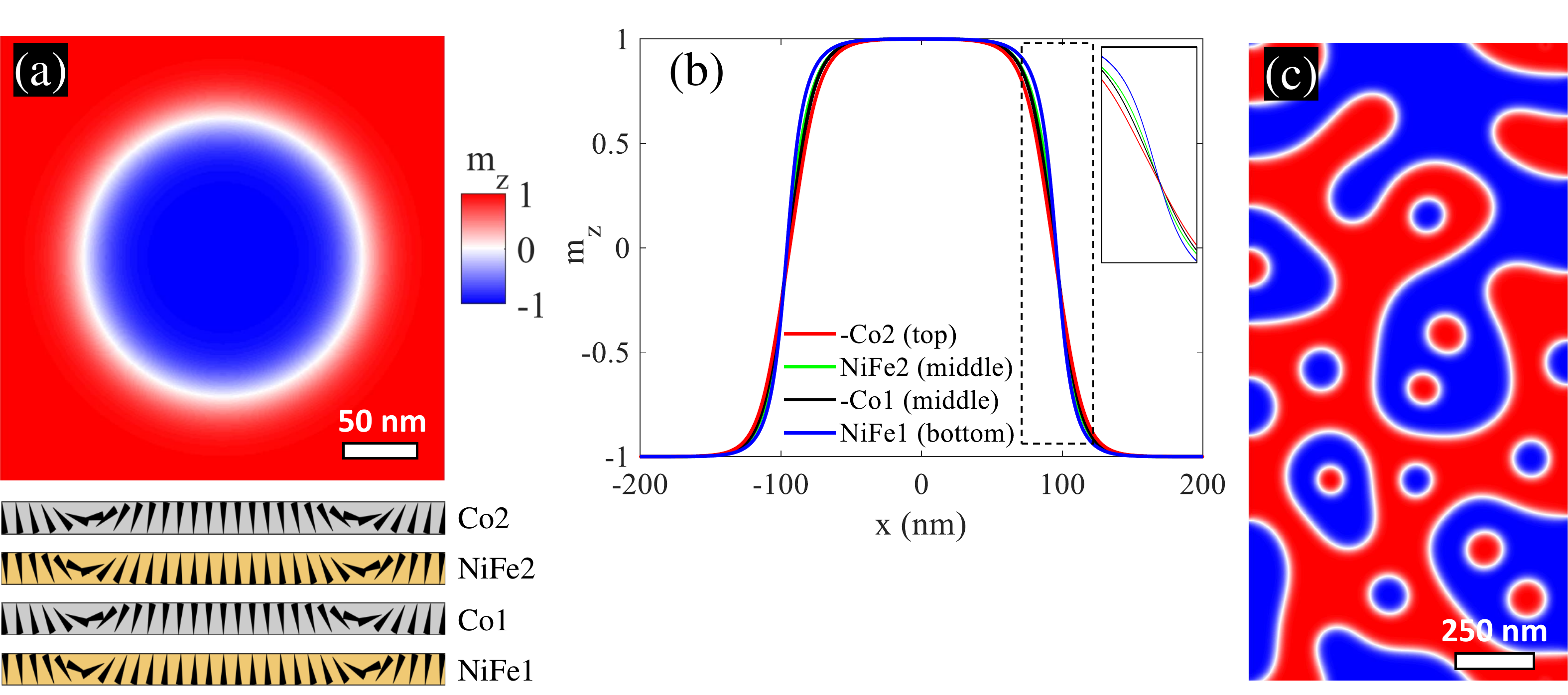}
\centering
\caption{\textbf{Micromagnetic simulation of skyrmions in SAF $\mid$} (a) (top) Spin texture of a skyrmion in the Co1 layer (m$_z$ in color scale). (bottom) Spin texture across the skyrmion diameter  within the different layers. (b) Out-of-plane component of the magnetization m$_z$ across the skyrmion diameter in the different layers. (c) Simulated spin texture in a $4.3\times8.5$ {\textmu}m$^2$ stripe.}
\label{simu}
\end{figure*}

\subsection*{Micromagnetic simulations and analytical modelling}

These experiments demonstrate  stable room-temperature skyrmions in compensated SAFs at zero external magnetic field. However, they raise several questions.  First, in these STXM experiments, only the out-of-plane component of the magnetization is accessible so that  the chirality of the spin texture cannot be determined. While left-handed N\'eel skyrmions are expected from the sign of the DMI at the Pt{\slash}Co interface~\cite{boulleRoomtemperatureChiralMagnetic2016}, the multilayer structure of the SAF leads to additional competing interlayer interactions which can affect the  skyrmion texture. In FM stacks, this interlayer interaction was shown to lead to twisted spin textures along the film thickness with the formation of N\'eel caps and layer-dependent chirality~\cite{legrandHybridChiralDomain2018,dovzhenkoMagnetostaticTwistsRoomtemperature2018}. While it is anticipated  that this effect would vanish in compensated SAF multilayers due to the  overall vanishing magnetic moment, the dipolar interaction between neighboring AF-coupled layers may also lead  to opposite chiralities between neighboring layers~\cite{hrabecCurrentinducedSkyrmionGeneration2017}. Second, the vanishing stray fields in SAFs are also expected to produce very small skyrmions, in the tens of nm range~\cite{buettnerTopologicalMassMagnetic2013}, while only large diameters are observed experimentally~\cite{legrandRoomtemperatureStabilizationAntiferromagnetic2019}.

To address these questions, we carried out micromagnetic simulations using experimentally extracted parameters (see Supplementary Information, S1.1). Fig.~\ref{simu}(a) shows a top view of the skyrmion spin texture in the middle Co layer (Co1) as well as a side view of the magnetization across the skyrmion in the different layers at zero field. The simulations confirm the existence of the SAF skyrmion solution in our system and show that the skyrmions in the different constituent layers all have uniform, left-handed chirality, in agreement with the sign of the DMI arising from the Pt{\slash}Co interface. The SAF skyrmions seen experimentally are hence homochiral, the chirality being driven by the DMI. This is also confirmed by additional magnetic microscopy (XMCD-PEEM) observations in similar stacks, where a direct observation of the skyrmion chirality of the top layer was achieved (see Supplementary Information, S2). Line-scans  along the skyrmion diameter provide the skyrmion profile within the different layers (see Fig.~\ref{simu}(b)): the profiles overlap in the domain but a slight misalignment is observed at the domain wall position. This can be accounted for by several effects related to the stray field energy: (i) a change of the domain wall width of the top (Co2) and bottom (NiFe1) layers induced by the uncompensated in-plane component of interlayer stray fields~\cite{hrabecMeasuringTailoringDzyaloshinskiiMoriya2014}   and  (ii) a lateral shift of the domain walls~\cite{baruth_domain_2006,hellwigDomainStructureMagnetization2007}. Both effects lead to non-zero local magnetic moments,  which allows to decrease the stray field energy at the expense of the interlayer exchange interaction.  Additionally, variations of the magnetic parameters (anisotropy and DMI) between the layers can affects the domain wall width.   Note that this lateral shift (less than 10 nm) could not be observed experimentally due to  the limited spatial resolution of the STXM (30 nm).
Fig.~\ref{simu}(c) shows the domain structure on a larger scale obtained from the relaxation of an initially demagnetized state. The simulations reproduce the experimental stripe domain state with magnetic skyrmions that is observed experimentally.

\begin{figure}[t!]
\includegraphics[width=0.55\textwidth]{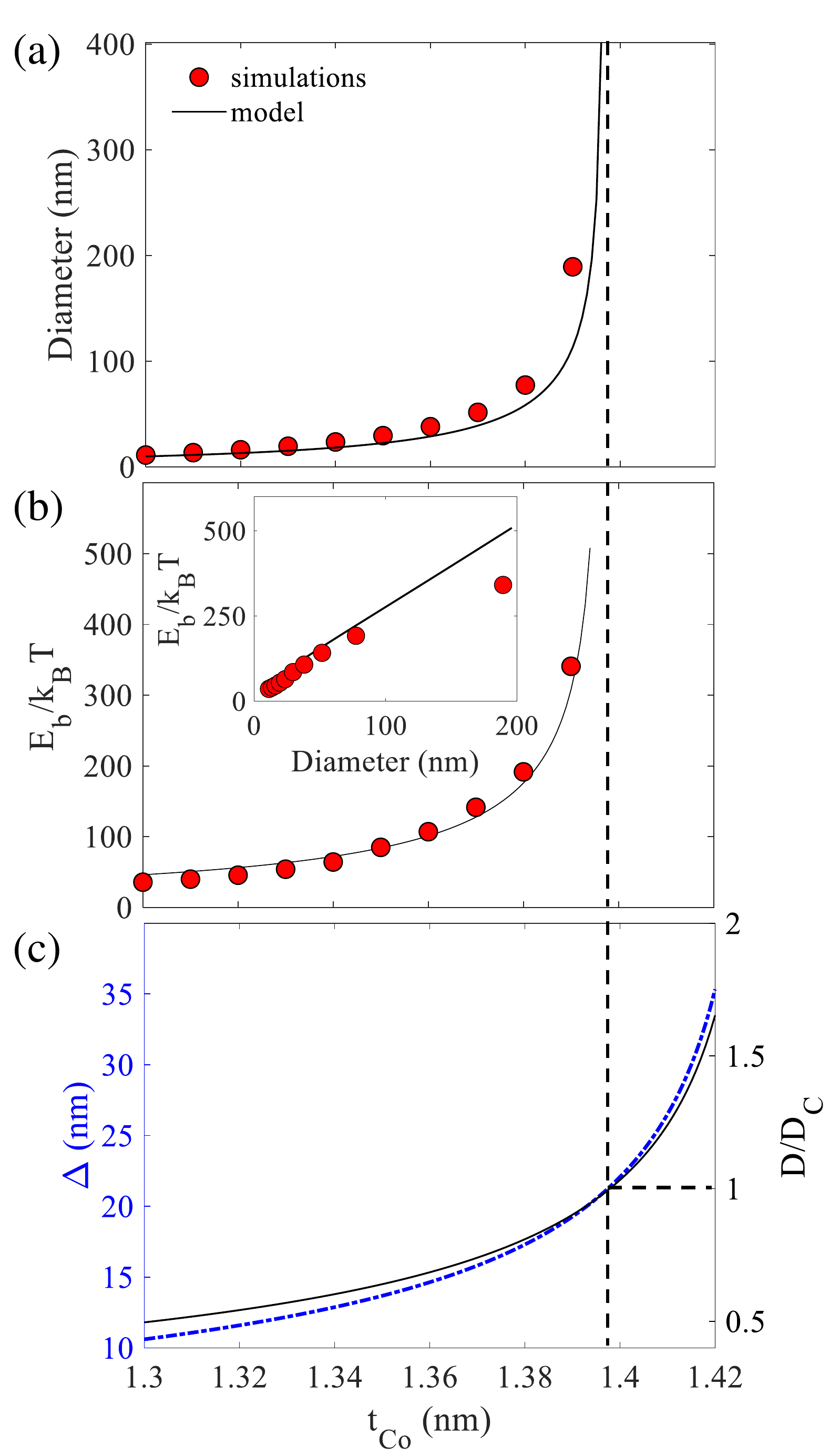}
\centering
\caption{\textbf{Analytical model $\mid$} (a) Skyrmion diameter  and (b) stability energy (b) vs the Co film thickness  predicted by the micromagnetic simulations (red dots) and  the analytical model (black line). The thickness of the Co{\slash}NiFe{\slash}Co layer is adjusted  such that the SAF is always compensated. Inset: stability energy vs skyrmion diameter.  (c) Domain wall width $\Delta$ and $D/D_c$ as a function of the film thickness.}
\label{model}
\end{figure}

The magnetic parameters  controlling the size and stability of the skyrmions in these ultra-thin SAF multilayers   depend on the film thickness of the different layers, in particular the interfacial PMA and DMI. This provides a way to  tune the skyrmion size and stability by playing on the film thickness.  Fig.~\ref{model}(a-b) show the skyrmion diameter and stability energy  $E_b$  as a function of the Co film thickness  as obtained from the simulations (T=0K, red dots). The  energy $E_b$ is defined as  $E_b=E_0-E_t$, where $E_t$ is the skyrmion energy with respect to the uniform state  as obtained from the micromagnetic simulations,  and $E_0=\sum_i 8\pi A_i t_i $  is the zero radius limit of the exchange energy in the continuous limit~\cite{Belavin_1975,buttnerTheoryIsolatedMagnetic2018,bernand-mantelSkyrmionbubbleTransitionFerromagnetic2018,leonovPropertiesIsolatedChiral2016} ($A_i$ and $t_i$ are the exchange constant and thickness of the $i^{th}$ layer respectively). Note that the thickness of the Pt{\slash}Co{\slash}NiFe{\slash}Co layer is adjusted accordingly such that the SAF is always compensated. The skyrmion diameter increases rapidly over a narrow range of film thickness,  around  1.39 nm corresponding to our experiments.  A similar trend occurs for the skyrmion stability energy. This is explained  by the linear relationship between the skyrmion size and its energy  (see inset of Fig. \ref{model}(b)): the larger the skyrmion, the more stable it is. Above 1.4 nm, no stable skyrmion state is obtained in the simulations. 

To better understand these results, we have built an analytical model to describe the size and stability of zero-magnetization skyrmions~\cite{ranno_design_2021}. This leads to the following expressions for the skyrmion radius $r_0$ and the skyrmion  energy $E$: 

\begin{align}
r_0 &=1.35 \Delta \frac{(\frac{D}{D_c})^2}{\sqrt{1-(\frac{D}{D_c})^2}} \label{eq1} \\
E &=8\pi A t(1-\frac{r_0}{4\Delta}) \label{eq2}
\end{align}

Here $t$ is the total FM film thickness (assuming identical materials for the two FM layers), $\Delta=\sqrt{A/K_{eff}}$ is the domain wall width, $D$ is the  DMI and $D_c=4\sqrt{AK_{eff}}/\pi$ is the critical DMI for which the single-domain state is no longer stable, $A$ is the exchange constant, $K_{eff}$ the effective perpendicular magnetic anisotropy. 
The dependence of the skyrmion diameter and stability energy on the film thickness is plotted in Fig.~\ref{model}(a-b), (black line) and a good agreement is obtained with the results of the micromagnetic simulation (red dots). Note that the model predicts a  linear relationship between the skyrmion radius and its stability energy, as observed in the  simulations. It also underlines that reducing the domain wall width enhances significantly the skyrmion stability for a given size, a result in agreement with  recent atomistic spin dynamics calculations~\cite{jia_material_2020}.   The model also reproduces well the divergence of the skyrmion diameter and stability energy when approaching a threshold value around 1.4 nm. This divergence is explained  by equation (\ref{eq1}): in this thickness range, $K_{eff}$  approaches zero (spin reorientation transition)  when the thickness increases. This first leads to a divergence of the domain wall width $\Delta$ (see Fig. \ref{model}) and second to $D/D_c$ approaching 1, \textit{i.e.} the domain wall energy tending toward zero. Both these effects lead to a fast increase of the skyrmion radius  upon  increasing the film thickness.

These results show that the large skyrmion size in this SAF stack can be accounted for by the low domain wall energy and large domain wall width resulting from the low perpendicular anisotropy.  Nevertheless,  the skyrmion diameter can be decreased down to 10 nm   by playing on the thickness of the layer constituting the SAF, while maintaining a significant thermal stability ($\sim 25 k_BT$). Note that here we considered two repetitions of the SAF bilayer (\textit{i.e} 4 FM layers), but the stability can  be further enhanced by increasing the number of repetitions since it is linear with the magnetic thickness in the limit of ultra-thin films. 

\begin{figure*}[t!]
\includegraphics[width=1\textwidth]{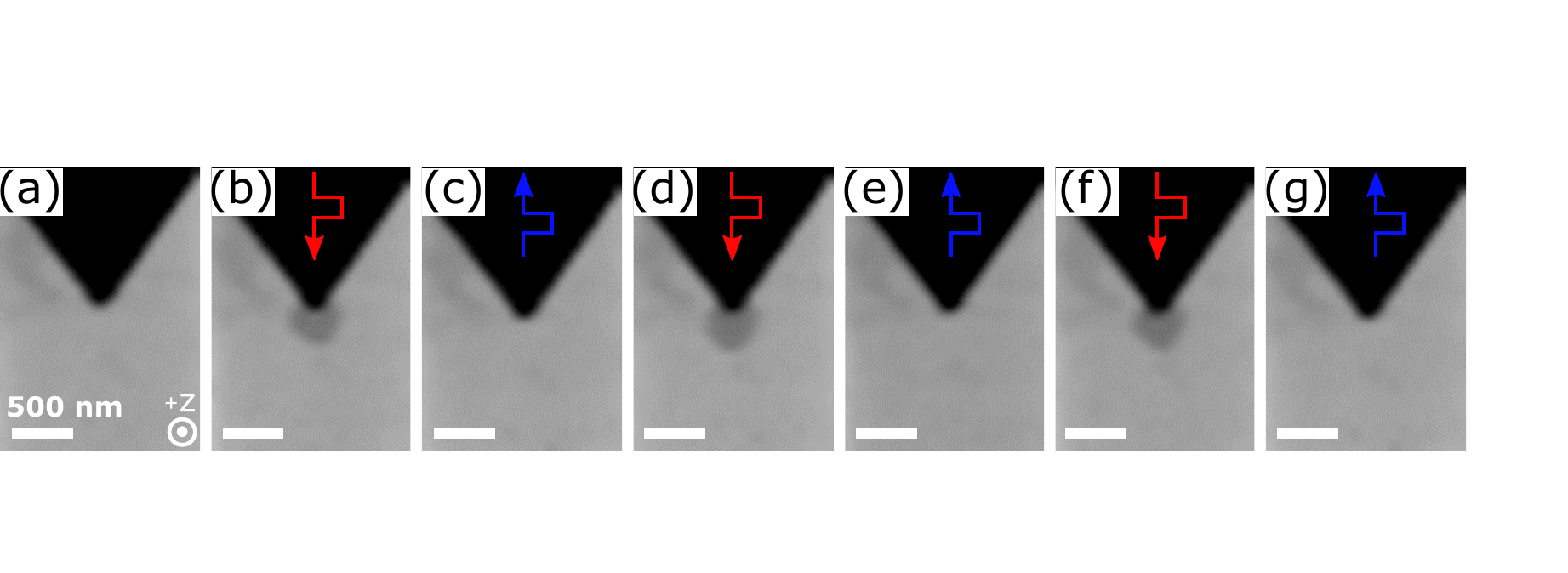}
\centering
\caption{\textbf{Current-induced nucleation{\slash}annihilation of SAF skyrmions $\mid$}  (a-g) Sequence of STXM images acquired at the Co edge. Before each acquisition, a single 5 ns current pulse with density $J=6.2\times10^{11}$ A/m$^2$ is injected in the direction indicated by the red and blue arrows. The layer composition is   [Pt{\slash}FM$_1${\slash}Ru{\slash}Pt{\slash}FM$_2${\slash}Ru]$_{12}$ with FM$_1$ = Co(0.2){\slash}NiFe(0.95){\slash}Co(0.2) and FM$_2$=Co(0.9) (thickness in nanometers). No external magnetic field is applied.}
\label{SAF_N=12_nucleation}
\end{figure*}

\subsection*{ Nucleation of SAF skyrmions using local current injection or ultrafast laser illumination}

In these experiments, the SAF skyrmion nucleation   using either external magnetic field or current injection  is stochastic and occurs on random sites in the track.
However, a local and controlled skyrmion nucleation is required for the write operation in devices.  Here, we show that this local and controlled skyrmion nucleation can be achieved using either pulsed current injection or ultrafast laser excitation at zero external magnetic field. 

For  the local current injection, a triangular shaped gold metallic tip  was patterned onto a 2-{\textmu}m-wide track (see STXM image in Fig.~\ref{SAF_N=12_nucleation}(a)): the large  current density at the apex of the triangle leads locally to large heating, spin accumulation,  and inhomogeneous current lines which can trigger the skyrmion nucleation under the tip~\cite{jiangBlowingMagneticSkyrmion2015,hrabecCurrentinducedSkyrmionGeneration2017,finizioDeterministicFieldFreeSkyrmion2019}. Here a SAF multilayer stack with a larger PMA was used so that  the tracks are uniformly magnetized at remanence (see Methods, SAF2). Starting from this single domain state at zero   field, the injection of a single 5 ns current pulse in the track  leads to the nucleation of a magnetic skyrmion under the tip (see   Fig.~\ref{SAF_N=12_nucleation}(b)). The skyrmions can then be annihilated by the subsequent injection of a current pulse with opposite polarity (Fig.~\ref{SAF_N=12_nucleation}(c)). This process is reproducible: the skyrmion can be nucleated and annihilated back and forth by the successive injection of current pulses of opposite polarities (Fig.~\ref{SAF_N=12_nucleation}(d-f)) and the nucleated skyrmion size remains fairly constant, around 420 nm. Furthermore, the process is found to be independent of the initial direction of the magnetization.  
Micromagnetic simulations using experimental parameters show that the skyrmion nucleation/annihilation process can be explained by the inhomogeneous current induced spin orbit torque under the metallic tip (see Supplementary Information).  
 
 \begin{figure*}[t!]
\includegraphics[width=0.7\textwidth]{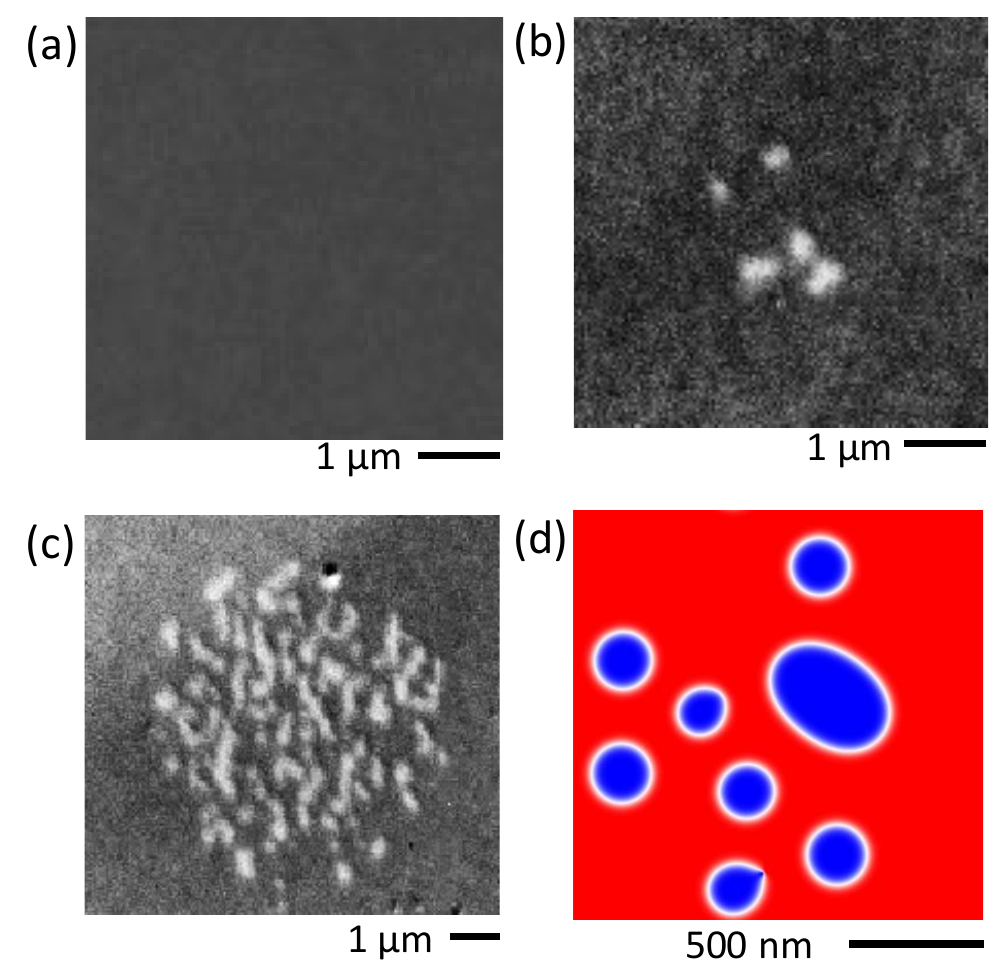}
\centering
\caption{\textbf{Nucleation of SAF skyrmions by ultrafast laser pulses $\mid$} (a-b) XMCD-PEEM images (Co $L_3$ edge) of a Pt(2.25){\slash}[Pt(0.75){\slash}Co(1.49){\slash}Ru(0.85)]$_{\times6}$ (thickness in nanometers) SAF multilayer (a) before and (b) during the illumination of the sample by ultrafast (100 fs) laser pulse excitation with repetition rate of 1.25  MHz and average power P=0.28 mW/{\textmu}m$^2$. (c) XMCD-PEEM image of worm  domains and skyrmions after the continuous illumination (several minutes) by   ultrafast laser pulses with repetition rate of 1.25 MHz and average power of 0.19 mW/{\textmu}m$^2$.  Experiments are carried out at zero external magnetic field. (d) Micromagnetic simulation using experimental parameters. Magnetization distribution (m$_z$ in color scale) obtained after relaxation  from a random initial state, mimicking a  fully demagnetized state resulting from the laser excitation. }
\label{laser}
\end{figure*}

Local nucleation of SAF skyrmions can also be achieved  using ultrafast laser pulse excitations. We show in Fig.~\ref{laser}(a)  an XMCD-PEEM image of a compensated SAF stack [Pt(0.75){\slash}Co(1.49){\slash}Ru(0.85)]$_{\times6}$ (see Methods, SAF3) at zero field exhibiting uniform magnetization. The thickness of the FM layer was adjusted such that the anisotropy is close to the spin reorientation transition so as to reduced the nucleation energy barrier, but the sample is still perpendicularly magnetized with large magnetic domains (see Supplementary Information, S1.3.2). The  illumination of the sample by  ultrafast (100 fs) laser pulses (repetition rate of 1.25 MHz) leads to the nucleation of several isolated SAF skyrmions at the laser spot position (Fig.~\ref{laser}(b)). Note that despite their vanishing magnetic moment, the SAF skyrmions are detectable with large magnetic contrast using XMCD-PEEM. This is explained by the PEEM surface sensitivity, which results in a larger contribution to the magnetic contrast   of the FM layers closer to the sample surface.  The nucleation of worm domains and isolated skyrmions on a larger area  by laser pulse trains (see (Fig.~\ref{laser}(c)) as well as single 100 fs laser pulses with larger  fluence, are also observed (see Supplementary Information, S1.3.2). The nucleation of magnetic skyrmions and skyrmion lattices by ultrafast laser pulses  was  demonstrated recently in  ultra-thin magnetic films~\cite{jeCreationMagneticSkyrmion2018,finazzi_laser-induced_2013,buttner_observation_2021} and is explained by  the laser-induced  heating which allows the skyrmion nucleation energy barrier to be overcome. To confirm this picture, we carried out micromagnetic simulations using experimentally derived parameters (see Supplementary Information, S1.3.3). To mimic the effect of the ultrafast demagnetization by laser, we considered as an initial state a magnetization distribution with  random orientation. After relaxation, SAF skyrmions are stabilized (see Fig. \ref{laser}(d)). Note that, in contrast to FM skyrmions, no external magnetic field is needed to nucleate SAF skyrmions from the uniformly magnetized state using ultrafast laser pulses.

\subsection*{Conclusion}

To conclude, we have demonstrated that skyrmions can be stabilized  at room temperature and zero magnetic field in compensated synthetic antiferromagnets.  Micromagnetic simulations and X-ray microscopy experiments confirm the left-handed N\'eel character of these SAF spin textures. An analytical model allows the identification of the physical parameters controlling the SAF skyrmion size and stability, which can be easily tuned by playing on the thickness of the different constituent layers. Since the skyrmion nucleation in SAFs is challenging due to their vanishing magnetic moment, we then studied the potential of local excitations. First, by designing injection tips, we showed the current-induced controlled and reversible nucleation and annihilation of skyrmions in an otherwise uniformly magnetized track. Second,  we demonstrate that isolated SAF skyrmions can be nucleated from a uniform magnetized state by 100 fs laser pulse excitation at zero magnetic field. The possibility to locally nucleate skyrmions by using current injection or ultrafast laser excitations in compensated SAFs offers  promising perspectives for devices based on the manipulation of SAF skyrmions.

\subsection*{METHODS}

\textbf{Sample preparation.} Three different compensated SAF stacks are discussed in the main text. They were designed with slightly different compositions to obtain optimal observation conditions in each experiment, depending on the observation and nucleation methods. Although their composition was slightly different, the samples all have very similar magnetic properties (see Supplementary Information) and bear the same underlying physics, making the current- and laser-induced nucleation mechanisms applicable to any of the three systems. The SAFs stacks are as follows: (i) \textbf{SAF1}, composed of Pt(2.5){\slash}[Pt(0.5){\slash}Co(1.35){\slash}Ru(0.85){\slash}Pt(0.5){\slash}Co(0.3){\slash}Ni$_{80}$Fe$_{20}$(1.45){\slash}Co(0.3){\slash}Ru(0.85)]$_{\times2}${\slash}Pt(2) (thickness in nanometers), was optimized for the observation of SAF skyrmions  by STXM (Fig. \ref{FIG_SAF_sk_N=2}), made possible by the presence of different FM materials in the SAF; (ii) \textbf{SAF2}, composed of Ta(3){\slash}Pt(3){\slash}[Pt(0.5){\slash}Co(0.2){\slash}Ni$_{80}$Fe$_{20}$(0.95){\slash}Co(0.2){\slash}Ru(0.85){\slash}Pt(0.5){\slash}Co(1.35){\slash}Ru(0.85)]$_{\times12}${\slash}Pt(2), was optimized with a slightly larger PMA, so as to facilitate the observation of the current-induced skyrmion nucleation and annihilation by starting from a uniformly magnetized film (Fig. \ref{SAF_N=12_nucleation}); (iii) \textbf{SAF3}, composed of Ta(3){\slash}Pt(2.25){\slash}[Pt(0.75){\slash}Co(1.49){\slash}Ru(0.85)]$_{\times6}${\slash}Pt(1.2) was used for the laser-induced skyrmion nucleation experiments, observed by XMCD-PEEM (Fig. \ref{laser}). SAF3 was kept close to the spin reorientation transition so as to minimize the skyrmion nucleation energy barrier. In addition, using surface-sensitive XMCD-PEEM allowed to use a stack composed of only one FM material (Co), simpler than SAF1 and SAF2 that were optimized to observe the antiparallel alignement of the skyrmions in the different FM layers by STXM. SAF1 and SAF2 were deposited by DC magnetron sputtering on both a 100 mm  Si (100) substrate and 200-nm-thick $100\times100$ {\textmu}m$^2$ Si$_3$N$_4$ membranes for the STXM experiments. The NiFe layer was deposited as a wedge, such that its thickness varies linearly along the  100 mm wafer. The films deposited on the membrane were  patterned into 2-{\textmu}m-wide tracks and contacted by Ti(10nm){\slash}Au(100nm) pads using standard nanofabrication processes. On some samples,  a triangular-shaped injection tip, consisting of Ti(10nm){\slash}Au(100nm), was patterned at one end of the track for the skyrmion nucleation. SAF3 was deposited by DC magnetron sputtering on a Si wafer as well as on transparent MgO substrates for the laser excitation experiments. The FM Co layer was deposited as a wedge, with thickness varying from 1 to 1.8 nm along the 100 mm Si wafer.  More details regarding the optimization of the SAFs and their characterization can be found in the Supplementary Information. \\

\textbf{X-ray magnetic microscopy experiments.} Scanning transmission X-ray magnetic microscopy experiments were carried out at the Hermes beamline of the Soleil synchrotron, Saint-Aubin, France; at the Maxymus beamline of the BESSYII synchrotron, Berlin, Germany; at the PolLux (X07DA) beamline of the SLS synchrotron, Villigen, Switzerland. All experiments were carried out at room temperature.  The ptychographic reconstructions were carried out at the Hermes beamline of the Soleil synchrotron. For that purpose, the following procedure was used: the point detector classically used for STXM (here a photo-multiplier tube) has been replaced by a CMOS camera~\cite{arora_spatially_2017}. The whole data set is obtained by scanning the sample with a 53$\%$ overlap of the probe. It is then reconstructed using PyNX software~\cite{favre-nicolin_pynx_2020} with the Alternate Projection algorithm~\cite{marchesini_augmented_2013}.
XMCD-PEEM imaging experiments were carried out at the CIRCE beamline \cite{foersterCustomSampleEnvironments2016} of the ALBA synchrotron, Barcelona, Spain, and at the UE49-PGMa beamline of the BESSYII synchrotron, Berlin, Germany. All experiments were carried out at room temperature. The laser-induced nucleation of SAF skyrmions were carried out at the UE49-PGMa beamline \cite{novakovic-marinkovicStripesBubblesDeterministic2020}. Gaussian-shaped laser pulses were generated by a Femtolasers Scientific XL Ti:Sapphire oscillator with a central wavelength of 800 nm, a pulse duration of 100 fs (full width at half maximum) and  circular polarization. The size of the laser beam was estimated to be around $18\pm6$ {\textmu}m$^2$ from the demagnetization area under continuous laser illumination at low power, in line with previous experiments~\cite{buttner_thermal_2020,arora_spatially_2017}. \\
 
\textbf{Micromagnetic simulations.} The micromagnetic simulations were carried out using Mumax3 \cite{vansteenkisteDesignVerificationMuMax32014}. The  simulation and magnetic parameters used for the simulation are described in the Supplementary Information.

\subsection*{ACKNOWLEDGEMENT}

We thank Jordi Prat for his support at ALBA synchrotron, HZB for the allocation of synchrotron radiation beamtime. We acknowledge the support of the Agence Nationale de la Recherche, project ANR-17-CE24-0045 (SKYLOGIC), the support of the DARPA TEE program through grant MIPR\#{} HR0011831554 from the DOI, the support of ALBA synchrotron through the CALIPSOplus (Grant 730872) funding, and the support of the PolLux beamline financed by the German Ministerium f\"ur Bildung und Forschung through contracts 05K16WED and 05K19WE2.

\newpage

\section*{Supplementary information }

\renewcommand{\thesection}{S\arabic{section}}
\renewcommand{\thesubsection}{S\arabic{section}.\arabic{subsection}}
\renewcommand{\thesubsubsection}{S\arabic{section}.\arabic{subsection}.\arabic{subsubsection}}
\renewcommand{\thefigure}{S\arabic{figure}}

\section{Preparation and characterization of the SAF stacks}

Three different compensated SAF stacks are discussed in the main text, with slightly different compositions, as detailed in the Methods section: (i) \textbf{SAF1}, composed of Pt(2.5){\slash}[Pt(0.5){\slash}Co(1.35){\slash}Ru(0.85){\slash}Pt(0.5){\slash}Co(0.3){\slash}Ni$_{80}$Fe$_{20}$(1.45){\slash}Co(0.3){\slash}Ru(0.85)]$_{\times2}${\slash}Pt(2) (thickness in nanometers), was used for the observation of SAF skyrmions at zero external magnetic field by STXM (Fig. 1, main text); (ii) \textbf{SAF2}, composed of Ta(3){\slash}Pt(3){\slash}[Pt(0.5){\slash}Co(0.2){\slash}Ni$_{80}$Fe$_{20}$(0.95){\slash}Co(0.2){\slash}Ru(0.85){\slash}Pt(0.5){\slash}Co(1.35){\slash}Ru(0.85)]$_{\times12}${\slash}Pt(2), was used for the current-induced skyrmion nucleation and annihilation (Fig. 4, main text); (iii) \textbf{SAF3}, composed of Ta(3){\slash}Pt(2.25){\slash}[Pt(0.75){\slash}Co(1.49){\slash}Ru(0.85)]$_{\times6}${\slash}Pt(1.2) was used for the laser-induced skyrmion nucleation experiments, observed by XMCD-PEEM (Fig. 5, main text).

\subsection{[Pt{\slash}Co{\slash}Ru{\slash}Pt{\slash}Co{\slash}NiFe{\slash}Co{\slash}Ru]$_{\times2}$ SAF1 stack}
\subsubsection{Optimization of the magnetic properties}

\begin{figure}[h!]
\includegraphics[width=0.8\textwidth]{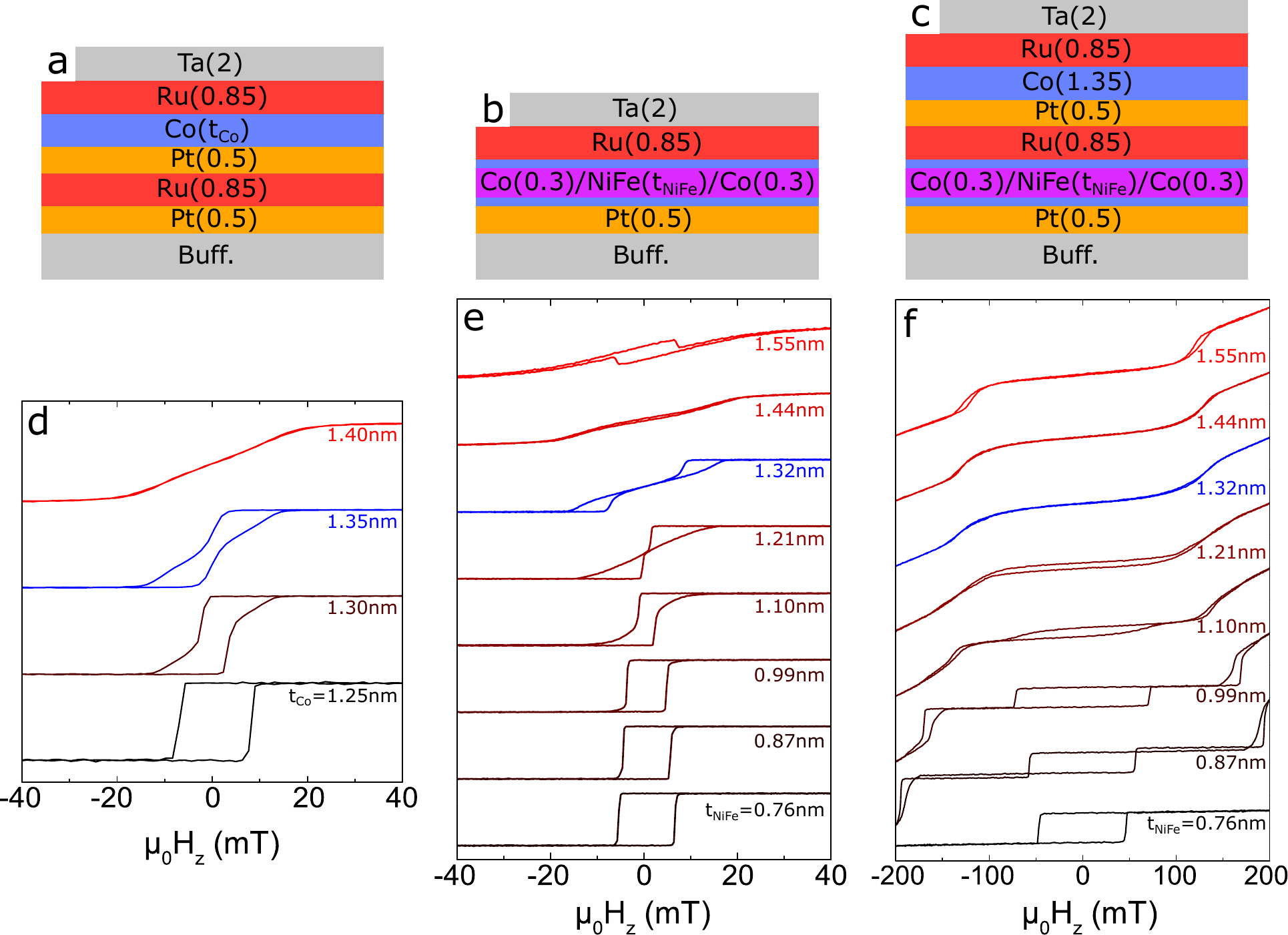}
\centering
\caption{\textbf{Optimization of the compensated stack SAF1 $\mid$} \textbf{a,b.} Material stacks for the constituent FM layers, \textbf{a.} Pt{\slash}FM$_2${\slash}Ru and \textbf{b.} Pt{\slash}FM$_1${\slash}Ru, with FM$_2$ = Co($t_{Co}$) and FM$_1$ = Co(0.3){\slash}Ni$_{80}$Fe$_{20}$($t_{NiFe}$){\slash}Co(0.3) (thickness in nanometers). \textbf{c.} Material stack for the SAF, Pt{\slash}FM$_1${\slash}Ru{\slash}Pt{\slash}FM$_2${\slash}Ru, with $t_{Co}=1.35$ nm. Buff. denotes Ta(3){\slash}Pt(2.5). The NiFe layer in \textbf{b} and \textbf{c} is deposited as a wedge. \textbf{d-f.} OOP MOKE hysteresis loops measured in the above stacks. The loops in \textbf{d} correspond to different (non-wedged) samples. In \textbf{e} and \textbf{f}, each loop is measured at positions 10 mm apart on the wedge.}
\label{FIG_loops_THICK}
\end{figure}

We discuss in this section the preparation and characterization of the  SAF1 sample corresponding to Fig. 1 in the main text.   The SAF1 stack   is  composed of a sputtered Pt{\slash}Co{\slash}NiFe{\slash}Co layer antiferromagnetically coupled to a Pt{\slash}Co layer through a thin  Ru layer to achieve RKKY AF coupling: [Pt(0.5){\slash}FM$_1${\slash}Ru(0.85){\slash}Pt(0.5){\slash}FM$_2${\slash} Ru(0.85)]$_{\times2}$ where FM1 = Co(0.3){\slash}Ni$_{80}$Fe$_{20}$(t$_{NiFe}$){\slash}Co(0.3) and FM$_2$ = Co(t$_{Co}$) (thickness in nanometers).  
The material stacks for the two FM layers, Pt{\slash}FM$_2${\slash}Ru and Pt{\slash}FM$_1${\slash}Ru, are represented in Fig. \ref{FIG_loops_THICK}.a and \ref{FIG_loops_THICK}.b respectively. In Fig. \ref{FIG_loops_THICK}.c the material stack for the SAF with $t_{Co}=1.35$ nm is represented. For each sample, the buffer layer (Buff.) consists of Ta(3){\slash}Pt(2.5).  Fig. \ref{FIG_loops_THICK}.d-f show the out-of-plane hysteresis loops measured by polar magneto-optical Kerr effect (pMOKE) in these three samples. In Fig. \ref{FIG_loops_THICK}.d, $t_{Co}$ is varied between 1.25 nm and 1.40 nm on different  samples and the loops display the gradual transition from a perpendicular magnetic anisotropy (PMA) to an in-plane  magnetic anisotropy (spin reorientation transition). In Fig. \ref{FIG_loops_THICK}.e, the Ni$_{80}$Fe$_{20}$ layer (hereafter NiFe) is deposited as a wedge with $t_{NiFe}=0.76-1.55$ nm. Here also, the loops display the gradual transition from PMA to IP (in-plane) anisotropy and, for $t_{NiFe}=1.32$ nm, a reversal characteristic of a multi-domain state is observed. 

Fig. \ref{FIG_loops_THICK}.f shows the loops for the SAF with $t_{Co}=1.35$ nm at different locations on the NiFe wedge. These loops exhibit two reversals, characteristic of antiferromagnetic RKKY coupling of perpendicularly magnetized layers. The amplitude of the central hysteresis, clearly visible for $t_{NiFe}\leq{}1.10$ nm, is proportional to the net magnetic moment. Upon increasing the NiFe thickness, this amplitude decreases and eventually vanishes for $t_{NiFe}\approx{}1.4$ nm, indicating magnetic moment compensation. This is accompanied by a widening of the central hysteresis, consistently with the diminution of the net magnetic moment. Note that the thickness of the Co layers surrounding the NiFe (0.3 nm) was chosen so that the spin reorientation transition of FM$_1$ matches that of FM$_2$. It also allows us to increase the magnetization of FM$_1$ to match that of FM$_2$, since $M_s\left(\mbox{NiFe}\right)<M_s\left(\mbox{Co}\right)$. The second reversal is followed by a constant-susceptibility region, corresponding to spin-flop processes, characteristic of SAFs with weak anisotropy \cite{dienyMagnetisationProcessesHysteresis1990, bloemenOscillatoryInterlayerExchange1994}. Note that the OOP saturation is not reached due to the limited applicable field in these experiments. The field at which this spin-flop transition occurs decreases with increasing NiFe thickness, as a result of the weakening of the anisotropy and of the RKKY coupling \cite{dienyMagnetisationProcessesHysteresis1990}.

The composition of SAF1 presented in the main text corresponds to $t_{Co}=1.35$~nm and $t_{NiFe}=1.45$~nm. The magnetic moment  of the different components of  SAF1 was measured by SQUID magnetometry:  a magnetic moment per unit area of $2.10\times10^{-3}\pm0.07$~A is measured for a Ta(3){\slash}Pt(3){\slash}Co(1.35){\slash}Ru(0.85){\slash}Pt(2) (thickness in nanometers) stack; a magnetic moment of $1.99\times10^{-3}\pm0.07$~A is measured for a Ta(3){\slash}Pt(3){\slash}Co(0.3){\slash}Ni$_{80}$Fe$_{20}$(1.45){\slash}Co(0.3){\slash}Ru(0.85){\slash}Pt(2) stack. 
Thus, the magnetic moment per area of the two ferromagnetic multilayers composing the stacks are  equal within error bars and the SAF stack is fully compensated. 
 
\subsubsection{Measurement of the Dzyaloshinskii-Moriya interaction by Brillouin light scattering spectroscopy}
\label{secSAF}

Brillouin light  scattering (BLS) spectroscopy experiments were carried out to extract the Dzyaloshinskii-Moriya interaction (DMI) of the Pt{\slash}Co{\slash}Ru and Pt{\slash}Co{\slash}NiFe{\slash}Co{\slash}Ru multilayers composing  SAF1. The principle of the measurement is the following~\cite{belmeguenaiInterfacialDzyaloshinskiiMoriyaInteraction2015,nembachLinearRelationHeisenberg2015, boulleRoomtemperatureChiralMagnetic2016,ranaRoomTemperatureSkyrmions2020}: the magnetization is saturated in the film plane by an external magnetic field and spin waves (SW) propagating along the direction perpendicular to this field are probed by a laser with a well-defined wave vector $k$ (Damon-Eshbach geometry). The DMI introduces a preferred chirality and  leads to an energy difference between  SW propagating with opposite wave vectors. This energy difference corresponds to a shift in frequency: $\Delta f (k) = f_S (k)-f_{AS} (k)$ where $f_S$ and $f_{AS}$ are the Stokes (a SW is created) and anti-Stokes (a SW is absorbed) frequencies respectively. This frequency shift is directly related to the DMI value $D$~\cite{belmeguenaiInterfacialDzyaloshinskiiMoriyaInteraction2015}: $\Delta f(k) = 2\gamma k D/(\pi M_s)$ with $\gamma=g \mu_B/\hbar$ the gyromagnetic ratio,  $M_s$ the saturation magnetization, $g$ the Land\'e factor. The DMI value $D$ can be extracted from the frequency-shift measured for both field polarities or from the slope of $\Delta_f$ for different $k$. To characterize the DMI in SAF1, we measured the DMI in the Pt{\slash}Co{\slash}Ru and Pt{\slash}Co{\slash}NiFe{\slash}Co{\slash}Ru stacks composing the SAFs. \\
 
\begin{figure*}[htb!]
\includegraphics[width=0.95\textwidth]{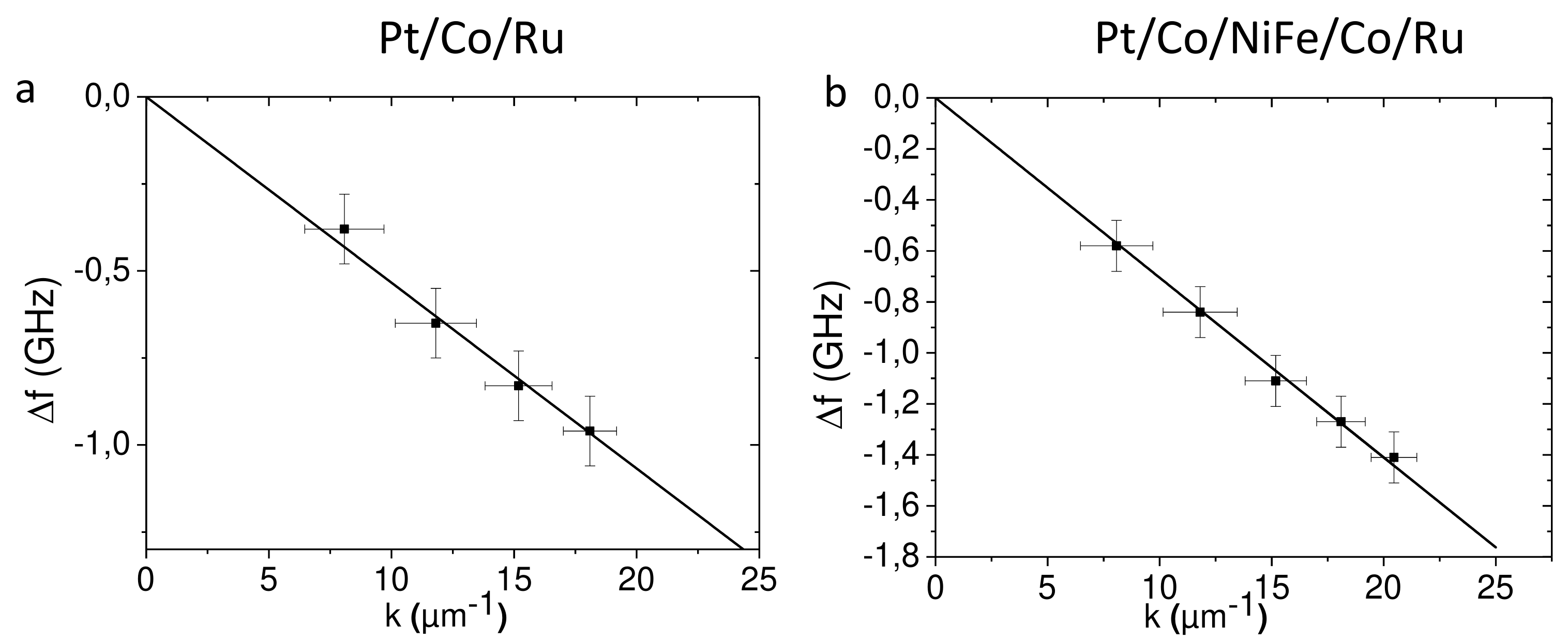}
\centering
\caption{\textbf{Brillouin light scattering measurements in SAF1 $\mid$} Frequency shift $\Delta f$ as a function of the spin wave vector $k$ for the sample \textbf{a.} Ta(3){\slash}Pt(3){\slash}Ru(0.85){\slash}Pt(0.5){\slash}Co(1.5){\slash}Ru(0.85){\slash}Pt(2) and \textbf{b.}   Ta(3){\slash}Pt(3){\slash}Co(0.3){\slash}NiFe(1.45){\slash}Co(0.3){\slash}Ru(0.85){\slash}Pt(2) (thickness in nanometers). The lines are linear fits assuming a zero intercept.}
\label{FigureBLS}
\end{figure*}

\paragraph{Pt{\slash}Co{\slash}Ru layers} 
The DMI was measured  using BLS in two   Pt{\slash}Co($t_{Co}$){\slash}Ru samples   with a Co thickness slightly above (1.5 nm) and slightly below (1.05 nm) the actual sample (1.35 nm). A first Pt{\slash}Co{\slash}Ru sample was measured by BLS with the  composition Ta(3){\slash}Pt(3){\slash}Ru(0.85){\slash}Pt(0.5){\slash}Co(1.5){\slash}Ru(0.85){\slash}Pt(2) (thickness in nanometers). The sample was deposited by magnetron sputtering on a Si(100) substrate.
Fig.~\ref{FigureBLS}.a shows $\Delta f$ as a function of the wave vector $k$.  The negative slope of $\Delta f$ vs $k$ indicates a  negative $D$  in agreement with our previous works~\cite{boulleRoomtemperatureChiralMagnetic2016}. From a linear fit of  $\Delta f$ vs $k$,   $D_{1.5}^{Co}=0.62\pm0.04$~mJ/m$^2$ is extracted. 
Here we used a  Land\'e factor $g=2.21$ as was measured by ferromagnetic resonance on a similar Pt{\slash}Co{\slash}MgO sample~\cite{jugeCurrentDrivenSkyrmionDynamics2019}. A saturation magnetization of $1.43\pm0.05\times10^6$~A/m was used as measured from the dependence of the magnetic moment on the ferromagnetic film thickness in Pt{\slash}Co{\slash}Ru thin films~\cite{bandieraAsymmetricInterfacialPerpendicular2011}. 

A second Pt{\slash}Co{\slash}Ru sample was measured by BLS with the same composition but with a thickness of Co of 1.05 nm. A frequency shift $\Delta f=6.22\times10^{-1}\pm1.97\times10^{-2}$~GHz is measured at  an in-plane wave vector $k_x=16.7$ \textmu{}m$^{-1}$ for both field polarities. 
This leads to a DMI $D_{1.05}^{Co}=0.86\pm0.06$~mJ/m$^2$. Assuming that the DMI scales linearly as $1/t$ between these two values,  $t$ being the film thickness, ~\cite{thiaville_dynamics_2012,belmeguenaiInterfacialDzyaloshinskiiMoriyaInteraction2015}, a DMI  $D_{1.35}^{Co}=0.67\pm0.04$~mJ/m$^2$ is estimated for a Co thickness of 1.35 nm corresponding to the Co layer of SAF1. \\

\paragraph{Pt{\slash}Co{\slash}NiFe{\slash}Co layers} 

The DMI was measured using BLS in  a sample with a composition close to the  Pt{\slash}Co{\slash}NiFe{\slash}Co{\slash}Ru layer in  SAF1.   The sample   with composition   Ta(3){\slash}Pt(3){\slash}Co(0.3){\slash}NiFe(1.45) {\slash}Co(0.3){\slash}Ru(0.85){\slash}Pt(2) (thickness in nanometers)  was deposited by magnetron sputtering on a Si(100) substrate.
Fig.~\ref{FigureBLS}.b shows $\Delta f$ as a function of the wave vector $k$ measured using BLS experiments.  From a linear fit of  $\Delta f$ vs $k$, $D=0.57\pm0.11$ mJ/m$^2$ is extracted.
Here we used an average  Land\'e factor of $g=2.13$  using $g=2.1$ for the NiFe layer~\cite{nembachLinearRelationHeisenberg2015} and 2.21 for the Co layer.
An average saturation magnetization of $0.97\pm0.15$~MA/m was estimated from the magnetic moment per area measured by SQUID magnetometry.

\subsubsection{Micromagnetic simulations}

The simulations were carried out with Mumax3 \cite{vansteenkisteDesignVerificationMuMax32014} using the parameters given in Table~\ref{TAB_param_simus}. These parameters correspond to those measured experimentally in the different layers composing the SAF stacks (see previous sections).

\begin{table}[h!]
\def\arraystretch{1.7}
\setlength{\tabcolsep}{3pt}
\centering
\begin{tabular}{|c|c|c|}
\hline
 & Co{\slash}NiFe{\slash}Co & Co  \\
\hline
$t_{eff}$ (nm) & 2.05 & 1.3916 \\
\hline
$M_s$ (MA/m) & 0.9707 & 1.43 \\
\hline
$K_u$ (MJ/m$^3$) & 0.6114 & 1.3206 \\
\hline
$D$ (mJ/m$^2$) & 0.57 & 0.67 \\
\hline
$A$ (pJ/m) & 6 & 16 \\
\hline
$\alpha$ & 0.08 & 0.12 \\
\hline
\end{tabular}
\caption{\textbf{Summary of the parameters $\mid$} Effective FM thickness, saturation magnetization, uniaxial anisotropy constant, DMI constant, exchange constant and magnetic damping.}
\label{TAB_param_simus}
\end{table}

Using the parameters in Table~\ref{TAB_param_simus}, we perform a hysteresis scan from positive field to negative field on a SAF sample of lateral dimensions $512\times512\rm nm^2$ with periodic boundary conditions along x and y. To obtain results in a reasonable time-frame, a bigger micromagnetic cell size of $4\times4\rm nm^2$ is assumed along the x-y dimension.

\begin{figure}[h!]
\includegraphics[width=0.5\textwidth]{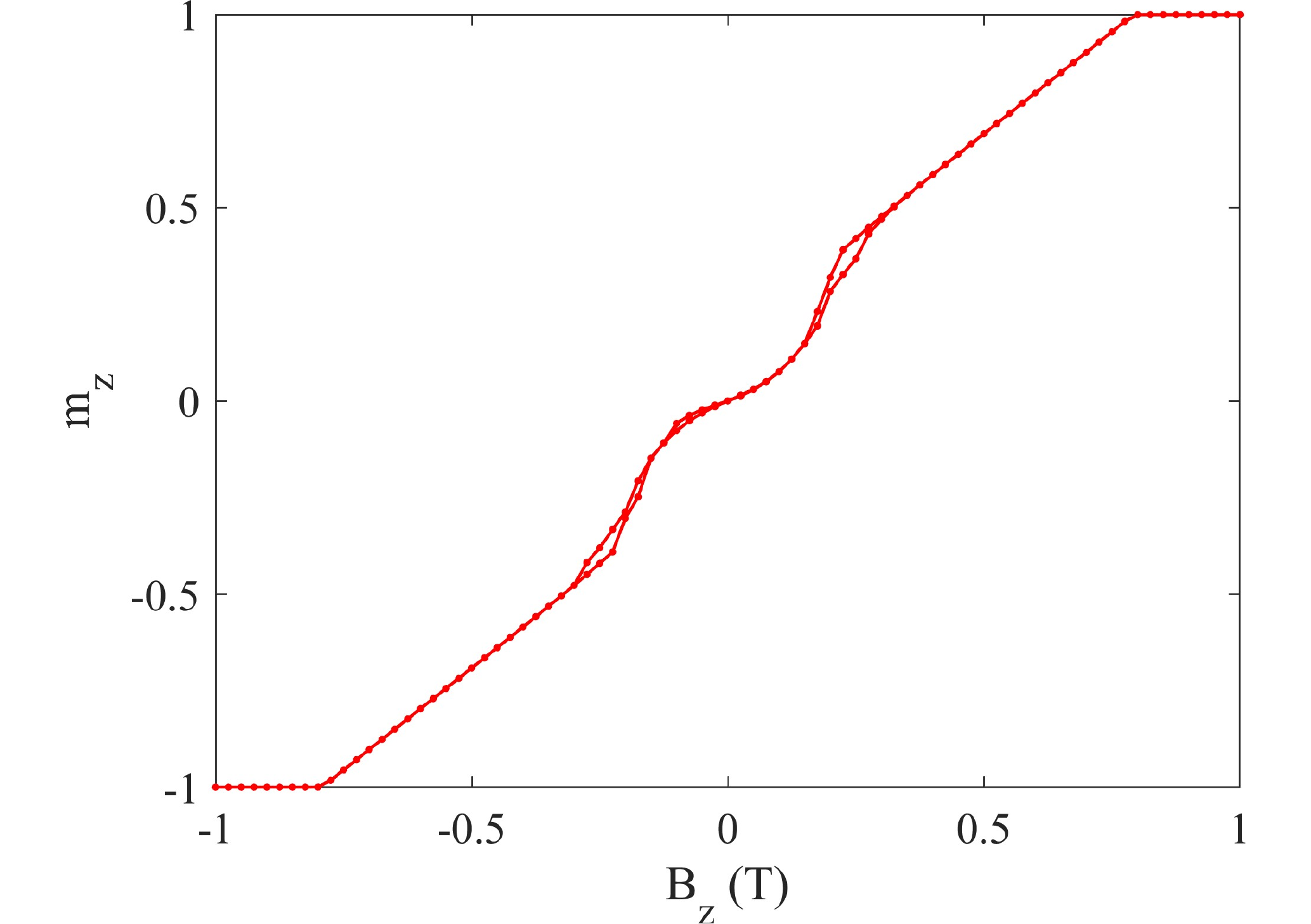}
\centering
\caption{Micromagnetic simulation of a hysteresis loop obtained for the parameters given in Table~\ref{TAB_param_simus}.}
\label{FIG_hyst_simu}
\end{figure}

\newpage

\subsection{[Pt{\slash}Co{\slash}Ru{\slash}Pt{\slash}Co{\slash}NiFe{\slash}Co{\slash}Ru]$_{\times12}$ SAF2 stack}
\subsubsection{Optimization of the magnetic properties}

\begin{figure}[h!]
\includegraphics[width=0.8\textwidth]{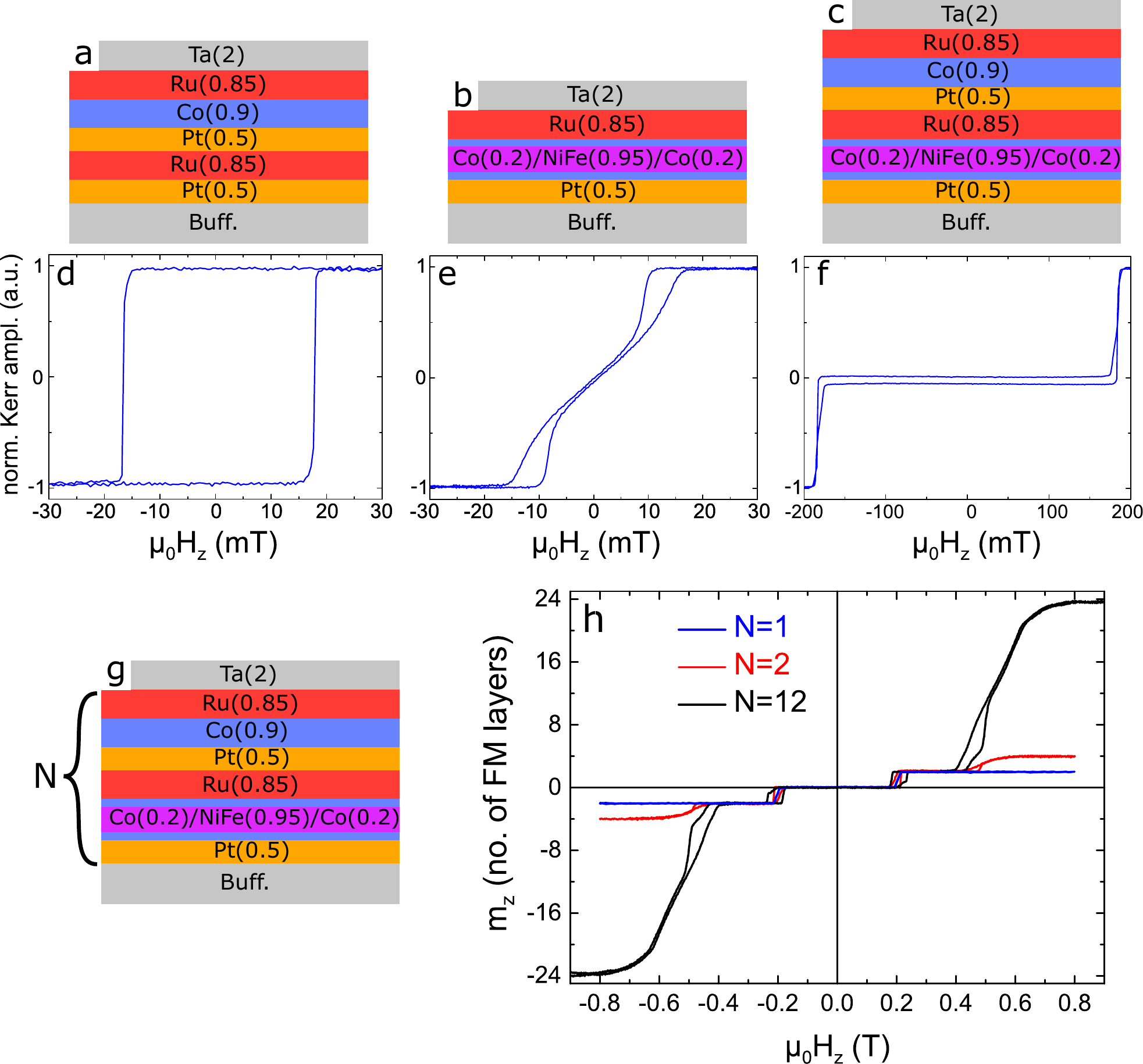}
\centering
\caption{\textbf{Compensated SAF multilayers with large PMA (SAF2) $\mid$} \textbf{a. b.} Material stacks for the constituent FM layers, \textbf{a.} Pt{\slash}FM$_2${\slash}Ru and \textbf{b.} Pt{\slash}FM$_1${\slash}Ru, with FM$_2$ = Co(0.9) and FM$_1$ = Co(0.2){\slash}Ni$_{80}$Fe$_{20}$(0.95){\slash}Co(0.2) (thickness in nanometers). \textbf{c.} Material stack for the SAF, Pt{\slash}FM$_1${\slash}Ru{\slash}Pt{\slash}FM$_2${\slash}Ru. Buff. denotes Ta(3){\slash}Pt(2.5). \textbf{d-f.} OOP MOKE hysteresis loops measured in the above stacks. \textbf{g.} Material stack for the multi-layered SAF. \textbf{h.} VSM hysteresis loops measured for different $N$. The signal is normalized and multiplied by the total number of constituent FM layers ($2N$).}
\label{FIG_loops_THIN}
\end{figure}

We discuss in this section the preparation and characterization of     sample SAF2  used for the demonstration of the current induced nucleation{\slash}annihilation of SAF skyrmion (see Fig. 4 in the main text). The SAF2 stack is composed of a sputtered Pt{\slash}Co{\slash}NiFe{\slash}Co layer antiferromagnetically coupled to a Pt{\slash}Co layer through a thin  Ru layer to achieve RKKY AF coupling: Ta(3){\slash}Pt(2.5){\slash}[Pt(0.5){\slash}FM$_1${\slash}Ru(0.85){\slash}Pt(0.5){\slash}FM$_2${\slash} Ru(0.85)]$_{\times12}${\slash}Pt(2) where FM1 = Co(0.2){\slash}Ni$_{80}$Fe$_{20}$(0.95){\slash}Co(0.2) and FM$_2$ = Co(0.9) (thickness in nanometers). The material stacks for the two constituent FM layers and for the SAF bilayer are shown in Fig. \ref{FIG_loops_THIN}.a, \ref{FIG_loops_THIN}.b and \ref{FIG_loops_THIN}.c respectively. Fig. \ref{FIG_loops_THIN}.d-f display the OOP hysteresis loops measured by magneto-optical Kerr effect (MOKE) in these three samples. Here, Pt{\slash}FM$_2${\slash}Ru exhibits full remanence due to the smaller Co thickness (0.9 nm), while Pt{\slash}FM$_1${\slash}Ru is kept close to the spin reorientation transition by reducing the thickness of the Co layers in contact with the NiFe to 0.2 nm. Fig. \ref{FIG_loops_THIN}.f shows the loop for the SAF bilayer ($N=1$)\footnote{The apparent net magnetic moment in Fig. \ref{FIG_loops_THIN}.f is probably an artefact due to the depth-sensitivity of the MOKE or because the maximum applicable field does not suffice to fully saturate the SAF. The vibrating sample magnetometry (VSM) loop (Fig. \ref{FIG_loops_THIN}.h, $N=1$), which provides a measurement of the total moment, confirms that the SAF is compensated.}. Fig. \ref{FIG_loops_THIN}.f reveals that the magnetization reversals are sharper, owing to the larger anisotropy.

For STXM experiments, we deposited SAF multilayers consisting of [Pt{\slash}FM$_1${\slash}Ru{\slash}Pt{\slash}FM$_2${\slash}Ru]$_N$ (Fig. \ref{FIG_loops_THIN}.g). The OOP hysteresis loops measured by VSM for different $N$ are shown in Fig. \ref{FIG_loops_THIN}.h, wherein the signal is expressed in units of the total number of constituent FM layers ($2N$). These loops exhibit a vanishing magnetic moment and zero susceptibility to the field in the central plateau region, which is characteristic of uniformly magnetized, compensated SAFs. For $N=1$, only one reversal exists, which defines the interlayer exchange field $\mu_0|H_{RKKY}|=200$ mT. For $N>1$, an additional reversal appears at $|H_z|=2H_{RKKY}$. It corresponds to the reversal of the internal FM layers, \textit{i.e.} all but the top-most and bottom-most FM layers: since each internal FM layer is AF-coupled on both sides (below and above), twice the external field is required to switch their magnetization \cite{hellwigDomainStructureMagnetization2007}. This can be seen from the amplitude of the different plateaus in Fig. \ref{FIG_loops_THIN}.h: for $H_{RKKY}<|H_z|<2H_{RKKY}$, the net magnetization is $|m_z|=2$, independently of $N$, while $|m_z|=2N$ at saturation ($|H_z|\gg{}2H_{RKKY}$). This means that $(N-1)$ layers have switched between the two plateaus.

\subsubsection{Measurement of the Dzyaloshinskii-Moriya interaction by Brillouin light scattering experiments}

To characterize the DMI in the SAF2 multilayer, we measured the DMI by BLS in Pt{\slash}Co{\slash}Ru and Pt{\slash}Co{\slash}NiFe{\slash}Co{\slash}Ru stacks with compositions close to the one of SAF2. 

\paragraph{Pt{\slash}Co{\slash}Ru layers} 

The DMI in the Pt(0.5nm){\slash}Co(0.9nm){\slash}Ru layer was estimated from the DMI measurement on the Ta{\slash}Pt{\slash}Ru{\slash}Pt(0.5nm){\slash}Co(1.05nm){\slash}Ru{\slash}Pt stack presented in section~S1.1.2. Assuming the DMI scales linearly as $1/t$, a DMI value $D=1.00\pm0.067$~mJ/m$^2$ is estimated. \\

\paragraph{Pt{\slash}Co{\slash}NiFe{\slash}Co layers} 

The DMI was measured by BLS in  a sample with a composition corresponding to the  Pt{\slash}Co{\slash}NiFe{\slash}Co{\slash}Ru layer in  SAF2. The sample   composition was Ta(3){\slash}Pt(3){\slash}Ru(0.85){\slash}Pt(0.5){\slash}Co(0.2){\slash}NiFe(0.96){\slash}Co(0.2){\slash}Ru(0.85){\slash}Pt(2) (thickness in nanometers) deposited by DC magnetron sputtering on a Si(100) substrate.A frequency shift $\Delta f=4.35\times10^{-1}\pm1.79\times10^{-2}$~GHz is measured at  an in-plane wave vector $k_x=16.7$ {\textmu}m$^{-1}$. This leads to a DMI $D=0.382\pm0.087$~mJ/m$^2$. Here we used an average Land\'e factor $g=2.13$ and an average saturation magnetization $M_s=0.88\pm0.16$ MA/m as measured by SQUID magnetometry.

\subsubsection{Observation of SAF skyrmions using Scanning transmission X-ray microscopy}

We then performed STXM experiments on this compensated SAF. The stack of Fig. \ref{FIG_loops_THIN}.g with $N=12$ was deposited on both a Si substrate (reference sample) and on a 200-nm-thick $100\times100$ {\textmu}m$^2$ Si$_3$N$_4$ membranes. The loop of Fig. \ref{FIG_loops_THIN}.h ($N=12$) was measured by VSM on the reference sample at the location of the membrane. The sample deposited on the membrane was then patterned into 2-{\textmu}m-wide tracks with Ti(10nm){\slash}Au(100nm) contact pads. In addition, a triangular-shaped injection tip, consisting of Ti(10nm){\slash}Au(100nm), was patterned at one end of the track for skyrmion nucleation.

\begin{figure}[h!]
\includegraphics[width=\textwidth]{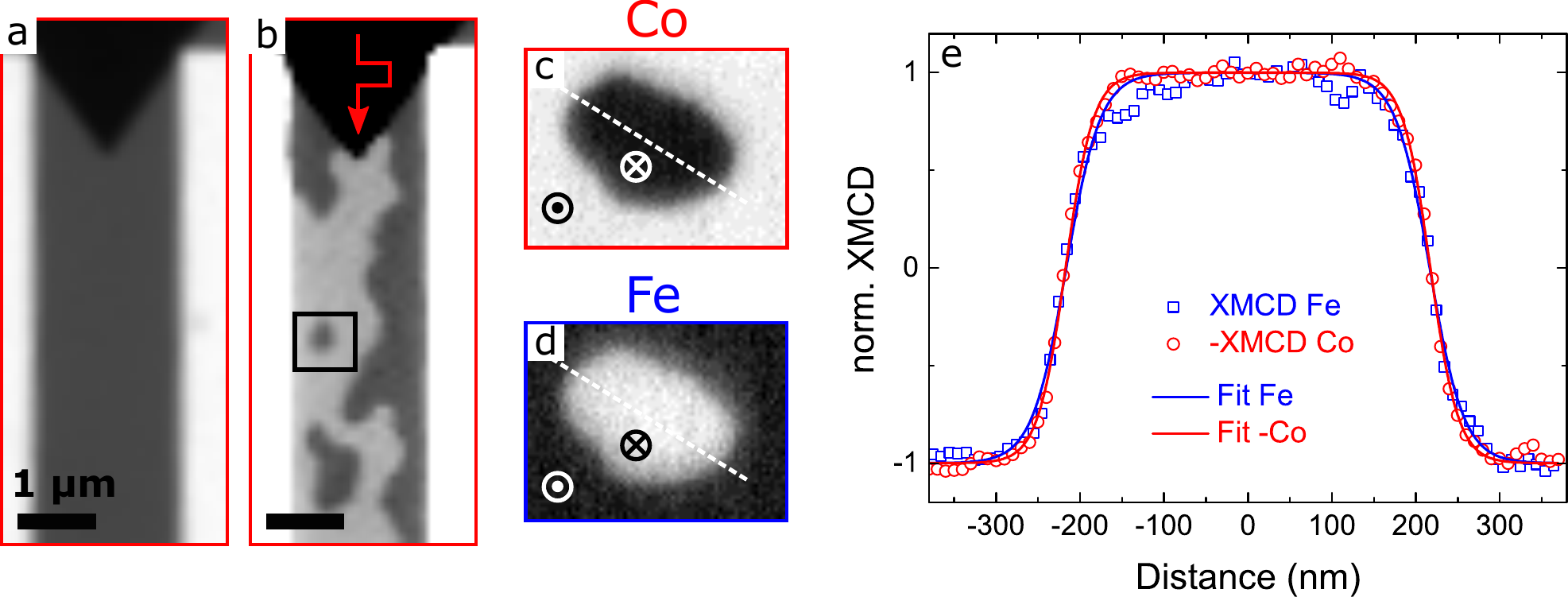}
\centering
\caption{\textbf{Observation of SAF skyrmions in SAF2 (large PMA) $\mid$} \textbf{a.} STXM image of a 2-{\textmu}m-wide at the Co edge. \textbf{b.} Same image after injection of multiple 10 ns current pulses of amplitude $J\approx{}10^{12}$ A/m$^2$. \textbf{c. d.} XMCD-STXM images acquired at \textbf{c.} the Co and \textbf{d.} the Fe edges in the area of \textbf{b} delimited by a rectangle. \textbf{e.} Normalized XMCD signal obtained from the line-scans along the white dashed lines in \textbf{c} and \textbf{d}. The signal for Co is inverted. The solid lines are fit with a 360{\textdegree} Bloch DW profile. All images were acquired at room temperature and zero magnetic field.}
\label{FIG_SAF_sk_N=12}
\end{figure}

Firstly, we image the magnetization texture. Fig. \ref{FIG_SAF_sk_N=12}.a shows a STXM image of a track acquired at the Co edge and at zero external magnetic field, that displays a uniformly magnetized state. Fig. \ref{FIG_SAF_sk_N=12}.b shows the same track after the injection of multiple 10 ns current pulses with $J\approx{}10^{12}$ A/m$^2$, leading to random nucleation of domains, most likely caused by the heating of the track. Fig. \ref{FIG_SAF_sk_N=12}.c and \ref{FIG_SAF_sk_N=12}.d show two XMCD-STXM images of a skyrmion acquired respectively at the Co and Fe absorption edges in the area marked with a rectangle in Fig. \ref{FIG_SAF_sk_N=12}.b. It points out the AF coupling between FM$_1$ and FM$_2$ as explained previously, which is emphasized in the plot of the XMCD contrast in Fig. \ref{FIG_SAF_sk_N=12}.e (open symbols). Here, due to the larger size of the skyrmion, the signal, proportional to $m_z$, is best fitted by a 360{\textdegree} Bloch DW profile (solid lines). Both signals superimpose  accurately, confirming the SAF character of the nucleated spin textures. Note that the effects observed in stray-field-coupled FM multilayers such as hybrid chiralities \cite{legrandHybridChiralDomain2018, dovzhenkoMagnetostaticTwistsRoomtemperature2018, liAnatomySkyrmionicTextures2019} are not expected is SAFs since the stray fields emanating from two adjacent FM layers compensate.

\begin{table}[h!]
\def\arraystretch{1.7}
\setlength{\tabcolsep}{3pt}
\centering
\begin{tabular}{|c|c|c|}
\hline
 & Co{\slash}NiFe{\slash}Co & Co  \\
\hline
$t_{eff}$ (nm) & 1.36 & 0.9 \\
\hline
$M_s$ (MA/m) & 0.9463 & 1.43 \\
\hline
$K_u$ (MJ/m$^3$) & 0.5948 & 1.6709 \\
\hline
$D$ (mJ/m$^2$) & 0.382 & 1 \\
\hline
$A$ (pJ/m) & 6 & 16 \\
\hline
$\alpha$ & 0.45 & 0.3 \\
\hline
\end{tabular}
\caption{\textbf{Summary of the parameters for nucleation/annihilation with current pulses $\mid$} Effective FM thickness, saturation magnetization, uniaxial anisotropy constant, DMI constant, exchange constant and magnetic damping.}
\label{TAB_param_simu_nuclea}
\end{table}

\subsubsection{Micromagnetic simulations of current induced nucleation of SAF skyrmions}
\begin{figure}[h!]
\includegraphics[width=0.8\textwidth]{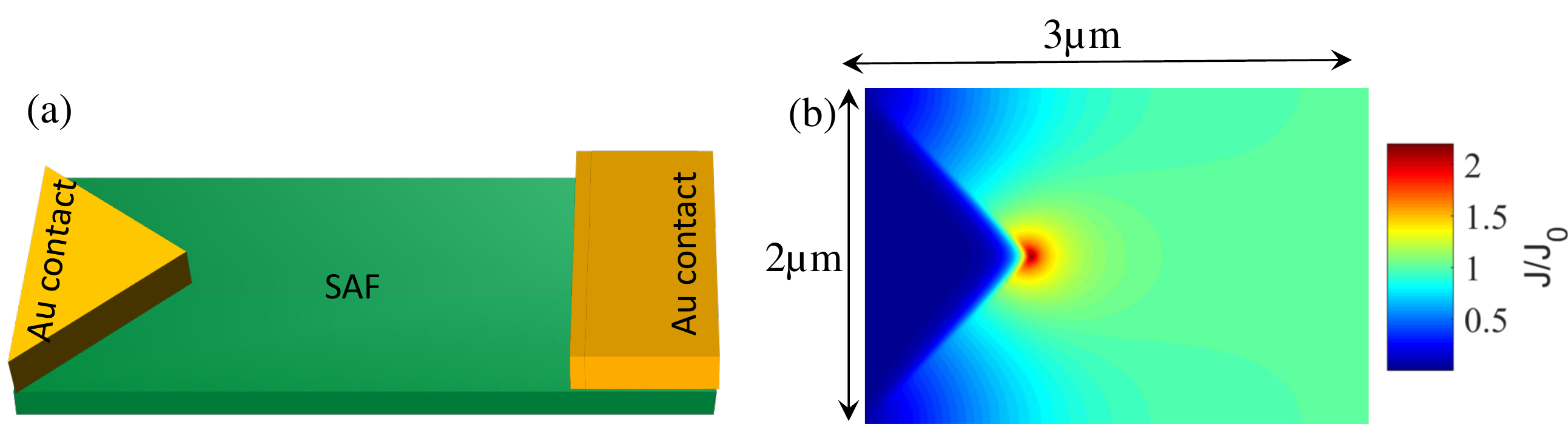}
\centering
\caption{\textbf{Current density simulations with COMSOL -} \textbf{a.} Schematic of simulated geometry in COMSOL. \textbf{b.} Current density map normalized w.r.t. amplitude far from the injection region}
\label{comsol}
\end{figure}

To elaborate on the results of nucleation/annihilation of skyrmions in SAF2 with current pulses, we perform corresponding numerical simulations. We first simulate a COMSOL model similar to the geometry used in experiments [Fig.~\ref{comsol}~\textbf{a.}]. We assume that the entire SAF structure has a homogeneous metallic composition with resistivity of 20 $\rm\mu \Omega.cm$. The resistivity value for the electrical contacts is 4 $\rm\mu \Omega.cm$ (for Au). The obtained current density is extracted on a finite difference grid and normalized w.r.t its amplitude at a position far from the current injection region, $J_0$ [Fig.~\ref{comsol}~\textbf{b.}]. This normalized current density map is then used as an input mask for further numerical simulations in $\rm mumax^3$. The input mask can be scaled back again inside $\rm mumax^3$ in proportion to the desired voltage pulse. The micromagnetic simulations are performed with only a single repetition of SAF with periodic boundary conditions along z-axis to mimic the N=12 SAF sample. The parameters are, however, consistent with the SAF2 sample as given in Table~\ref{TAB_param_simu_nuclea}. A slightly larger cell size of $\rm 2~nm \times 2~nm \times z~nm$ is used to simulate the large sample size ($\rm 3\mu m \times 2\mu m \times z~nm$). The simulation also contains randomly distributed grain structure with small variation (5\%) in $K_u$ and D parameters. The anisotropy axis of the grains is also randomly allowed to deviate away from z-axis up to a maximum of $4^\circ$. The inter-grain exchange is also reduced by 10\%. The simulations are performed at T=0~K.

\begin{figure}[h!]
\includegraphics[width=0.8\textwidth]{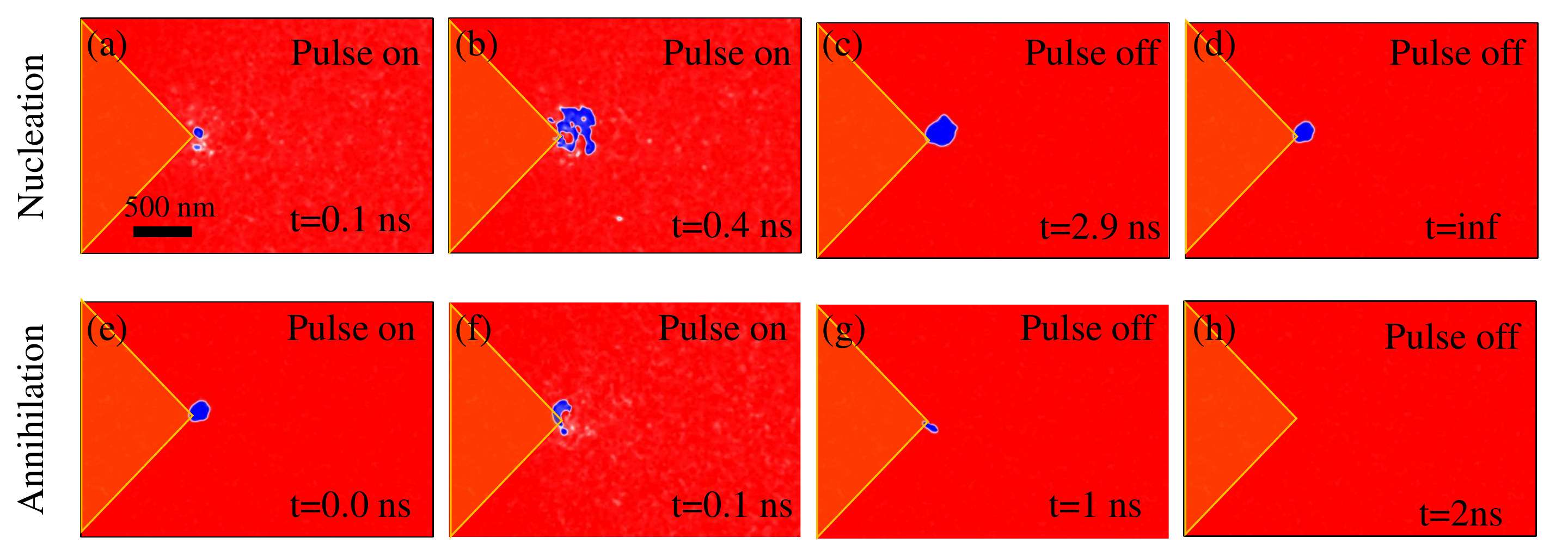}
\centering
\caption{\textbf{Micromagnetic Simulations - Current-induced nucleation/annihilation} \textbf{a.} Evolution of magnetization during skyrmion nucleation on application of current density pulse of $\rm J_0= 15\times10^{12}~A/m^2$ and width $\rm 0.4~ns$. \textbf{b.} Skyrmion annihilation process with opposite polarity pulse of $\rm J_0=- 15\times10^{12}~A/m^2$ and width $\rm 0.15~ns$}
\label{nucleation_sim}
\end{figure}

In Fig.~\ref{nucleation_sim}~\textbf{a-d}, we show the nucleation of a skyrmion using a current pulse of $\rm J_0= 15\times10^{12}~A/m^2$ and pulse width of $\rm 0.4~ns$. After the pulse, the generated texture is first allowed to evolve by integrating LLG followed by an energy minimization to establish the final stable state [Fig.~\ref{nucleation_sim}~\textbf{d.}]. We note here that the skyrmion in Fig.~\ref{nucleation_sim}~\textbf{d.} is only stabilized due to the pinning provided by the randomly distributed grains as the parameters used for SAF2 do not allow the skyrmion to be a stable state. We also find that for our simulations, no nucleation can occur below a threshold current density of $\rm J_0= 12\times10^{12}~A/m^2$. This current density is around twenty times larger than the current density used in the experiments. We attribute this mismatch to the contribution of Joule heating in the nucleation process. At the tip region, where the current density is largest, Joule heating can allow the nucleation process to begin at reduced threshold current densities. Since our simulations are performed at 0~K, an increase in current density required for nucleation is expected. We have confirmed this reduction in threshold current from similar simulations performed at elevated temperatures. However, an exact quantitative picture for thermal contributions in a micromagnetic model is hard to establish given the limitations of the Brown’s approach for inclusion of thermal field. In Fig.~\ref{nucleation_sim}~\textbf{e-h}, we also show the annihilation of the skyrmion nucleated in Fig.~\ref{nucleation_sim}~\textbf{d} using the same current pulse, $\rm J_0= 15\times10^{12}~A/m^2$, but with opposite polarity. The pulse width is $\rm t=0.15~ns$ after which the skyrmion slowly shrinks and subsequently annihilates.

\subsubsection{Local current induced nucleation of SAF skyrmionium}
\begin{figure}[h!]
\includegraphics[width=0.8\textwidth]{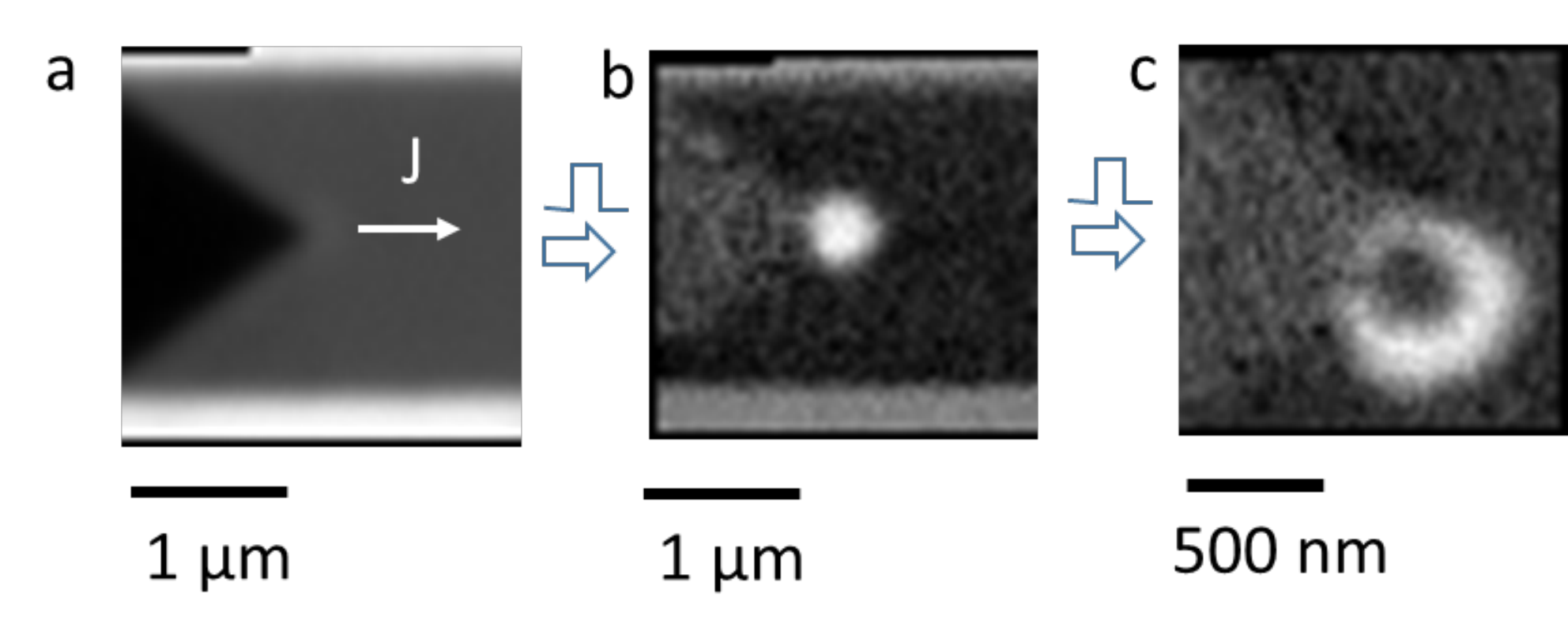}
\centering
\caption{\textbf{Nucleation of a skyrmionium spin texture $\mid$} \textbf{a.} XMCD-STXM image (Co $L_3$ edge) of the SAF magnetic wire with the triangular gold electrode for current injection. \textbf{b.} Same image but after the injection of a 5.3 ns current pulse with density 7$\times10^{11}$ A/m$^2$ and \textbf{c.} the subsequent injection of  a  5.2  ns current pulse with density $5.6\times10^{11}$ A/m$^2$ .}
\label{Skyrmionium}
\end{figure}

 The local current induced nucleation in the SAF2 stack was studied during   STXM beamtimes in Hermes beamline in Soleil Synchrotron, Saint-Aubin (France) in Maxymus beamline in Bessy synchrotron, Berlin (Germany), as well as in Pollux beamline in SLS synchrotron (Villigen, CH).  As shown in the main text, we have demonstrated the reproducible nucleation{\slash}annihilation of SAF skyrmions using current injection.  
We show in Fig.~\ref{Skyrmionium}  that the successive injection of two positive pulses also allows us to nucleate \textit{double skyrmion} spin structure referred to as skyrmionium.

\newpage

\subsection{(Pt{\slash}Co{\slash}Ru)$_{\times6}$ SAF3 stack}
 
\subsubsection{SAF3 composition and optimization}

\begin{figure}[h!]
\includegraphics[width=1\textwidth]{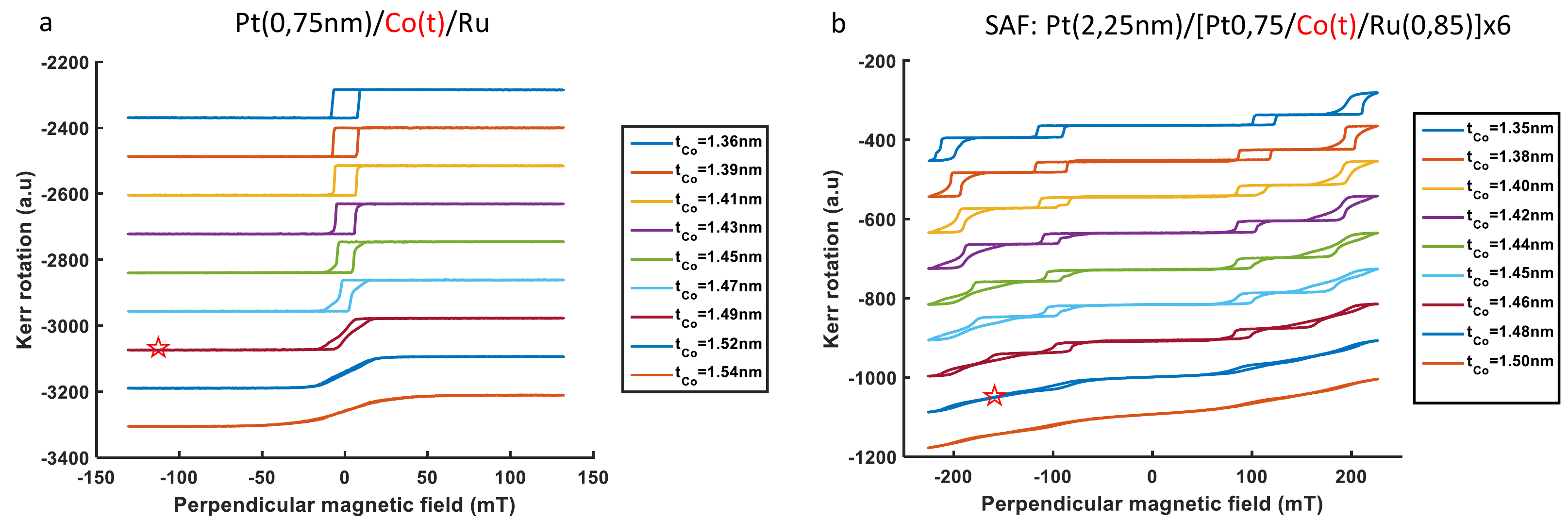}
\centering
\caption{\textbf{Optimization of SAF3 - Hysteresis loops $\mid$} Kerr rotation vs magnetic field applied perpendicularly to the sample plane for \textbf{a.} a Ta(3){\slash}Ru(0.85){\slash}Pt(0.75 nm){\slash}Co(t){\slash}Ru(0.85){\slash}Ta(1.5) and \textbf{b.}  Pt(2.25nm){\slash}[Pt(0.75){\slash}Co(t){\slash}Ru(0.85]x6 for different Co thicknesses corresponding to the same positions along the diameter of the 100 mm wafer (curves are offset for clarity).  The Co thicknesses varies between 1.33 nm and 1.55 nm. The red stars highlight the hysteresis loop of the sample where the ultrafast laser nucleation was observed.}
\label{FigureMaterialLaserSupp}
\end{figure}

The composition of the SAF3 multilayer stacks  used for the nucleation of the   skyrmions using laser excitation was the following: 
Ta(3){\slash}Pt(2.25){\slash}[Pt(0.75){\slash}Co(1.49){\slash} Ru(0.85)]$_{\times6}${\slash}Pt(1.2) (thickness in nanometers). The samples were deposited by magnetron sputtering on Si wafers as well as transparent MgO wafers for the laser excitation. The ferromagnetic Co layer was deposited using a wedge deposition so that the thickness of the Co layer was varying linearly along the  the 100 mm diameter Si wafer between 1 and 1.8 nm.
We show in Fig.~\ref{FigureMaterialLaserSupp}(a)  hysteresis loops of a   Ta(3){\slash}Pt(3){\slash}Ru(0.85){\slash}Pt(0.75nm){\slash}Co(t){\slash}Ru(0.85 nm){\slash}Ta(1.5) layer, which is the essential brick composing the SAF, at different positions along the diameter of the wafer close to its center. As the thickness decreases, the hysteresis loops gradually change from a square shape, indicating a perpendicular magnetization orientation, to a linear and reversible reversal, indicating an in-plane magnetization orientation. The transition occurs around a Co thickness of 1.5 nm.  The hysteresis loops of the corresponding SAF structures are shown in Fig. \ref{FigureMaterialLaserSupp}.b. As expected, a flat hysteresis loop with zero Kerr signal is observed at remanence, due to the full compensation of the SAF. 

\begin{figure*}[htb!]
\includegraphics[width=0.5\textwidth]{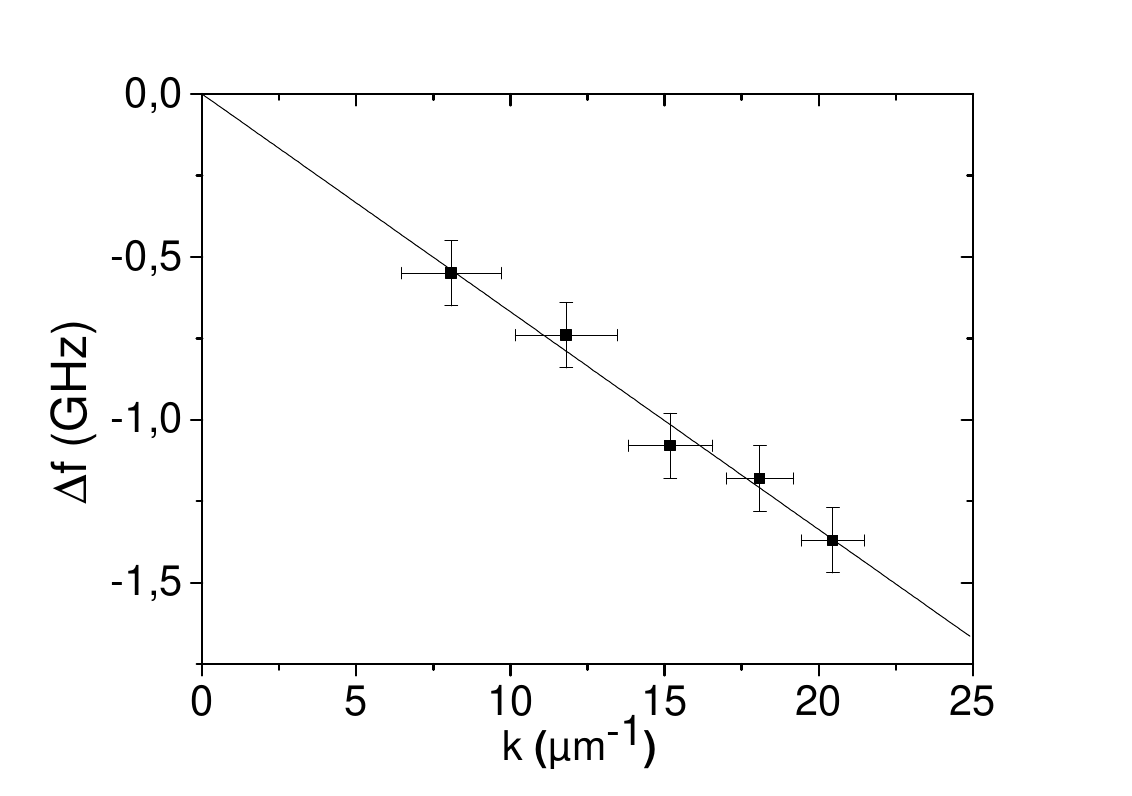}
\centering
\caption{\textbf{Brillouin light scattering measurements in SAF3 $\mid$} Frequency shift $\Delta f$ as a function of the spin wave vector $k$ for the sample a. Ta(3){\slash}Pt(3){\slash}Ru(0.85){\slash}Pt(0.75){\slash}Co(1.5){\slash}Ru(0.85){\slash}Pt(2) (thickness in nanometers).}
\label{FigureBLS2}
\end{figure*}

The DMI was measured  using BLS in a Ta(3){\slash}Pt(3){\slash}Ru(0.85){\slash}Pt(0.75){\slash}Co(1.5){\slash}Ru(0.85){\slash}Pt(2) (thickness in nanometers) deposited by magnetron sputtering on a Si(100) substrate.
Fig.~\ref{FigureBLS2}.a shows $\Delta f$ as a function of the wave vector $k$. From a linear fit of  $\Delta f$ vs $k$,   $D=0.7725\pm0.041$~mJ/m$^2$ is extracted. 
Here we used a  Land\'e factor $g=2.21$~\cite{jugeCurrentDrivenSkyrmionDynamics2019} and a saturation magnetization of $1.43\pm0.05\times10^6$~A/m  as measured from the dependence of the magnetic moment on the ferromagnetic film thickness in Pt{\slash}Co{\slash}Ru film~\cite{bandieraAsymmetricInterfacialPerpendicular2011}. 

\subsubsection{Nucleation of SAF skyrmions using single fs laser pulse excitations}
\begin{figure}[h!]
\includegraphics[width=1\textwidth]{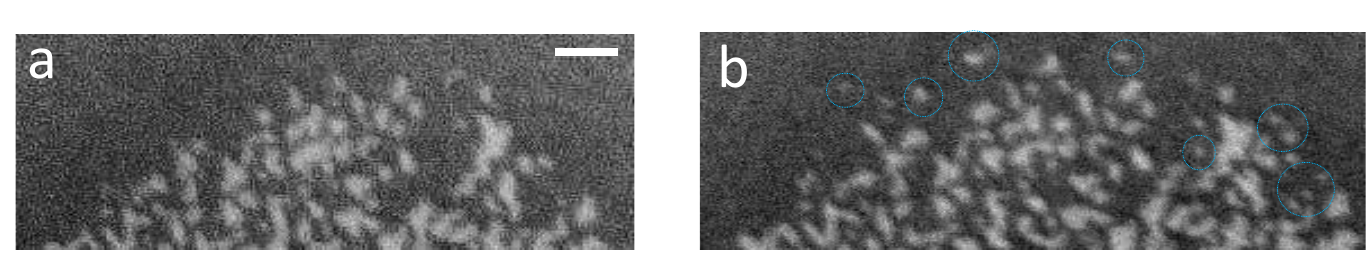}
\centering
\caption{\textbf{Nucleation of SAF skyrmions in SAF3 using single fs laser pulse excitation $\mid$} XMCD-PEEM images \textbf{a.} before and \textbf{b.} after the illumination of the sample by a single ultrafast (100 fs) laser pulse excitation with fluence 76  mJ/cm². The circles show the newly nucleated magnetic skyrmions. Experiments are carried out at zero external magnetic field. The scale bar is 1 {\textmu}m long.}
\label{Laser_Single_sup}
\end{figure}

To observe the magnetic skyrmions and nucleate them, we used the XMCD-PEEM magnetic microscope of the  PGM beamline UE49-PGMa beamline in Bessy synchrotron, Berlin (Germany) featuring an in-situ ultrafast laser excitation pulse. The sample had a Co thickness of 1.49 nm, where the anisotropy is close to the in-plane to out-of-plane reorientation transition (see Fig.~\ref{FigureMaterialLaserSupp}.a, red star), but the sample is still perpendicularly magnetized.  We show in Fig.~\ref{Laser_Single_sup} XMCD-PEEM images of isolated worm domains and skyrmions before (a) and after (b) the illumination of  the SAF sample with a single 100 fs laser pulse leading to the nucleation of SAF    skyrmions. The comparison of the two images shows that isolated SAF skyrmions were  nucleated by single 100 fs laser pulses  (see blue circles).

\subsubsection{Micromagnetic simulation of the laser induced nucleation}

Micromagnetic simulations were carried out to  characterize the laser induced nucleation of the SAF skyrmions. For simulations of laser induced nucleation, we use the following parameters extracted from experiments corresponding to Fig. 3 in the main paper. To mimic the thermal excitation by the laser pulse, we start our simulation with a random magnetization configuration in a circular region (laser spot) in the center of the uniformly magnetized SAF structure. The energy of this configuration is then minimized to achieve the final stable state. Note that although micromagnetic simulations capture most of the physics of the skyrmion formation after the laser-induced demagnetization on a longer time scale, atomistic spin dynamics calculations would be needed to describe the laser-induced demagnetization and the dynamics just after the laser pulse on the ps time scale. Note also that the simulations are done at 0K and  do  not take into account the temperature dependence of the magnetic parameters.

\begin{table}[h!]
\def\arraystretch{1.7}
\setlength{\tabcolsep}{3pt}
\centering
\begin{tabular}{|c|c|}
\hline
 & Co  \\
\hline
$t_{eff}$ (nm) & 1.49 \\
\hline
$M_s$ (MA/m) & 1.43 \\
\hline
$K_u$ (MJ/m$^3$) & 1.311 \\
\hline
$D$ (mJ/m$^2$) & 0.722 \\
\hline
$A$ (pJ/m) & 16 \\
\hline
\end{tabular}
\caption{\textbf{Summary of the parameters for Laser induced nucleation $\mid$} Effective FM thickness, saturation magnetization, uniaxial anisotropy constant, DMI constant and exchange constant.}
\label{TAB_param_simu_laser}
\end{table}

\begin{figure}[h!]
\includegraphics[width=0.75\textwidth]{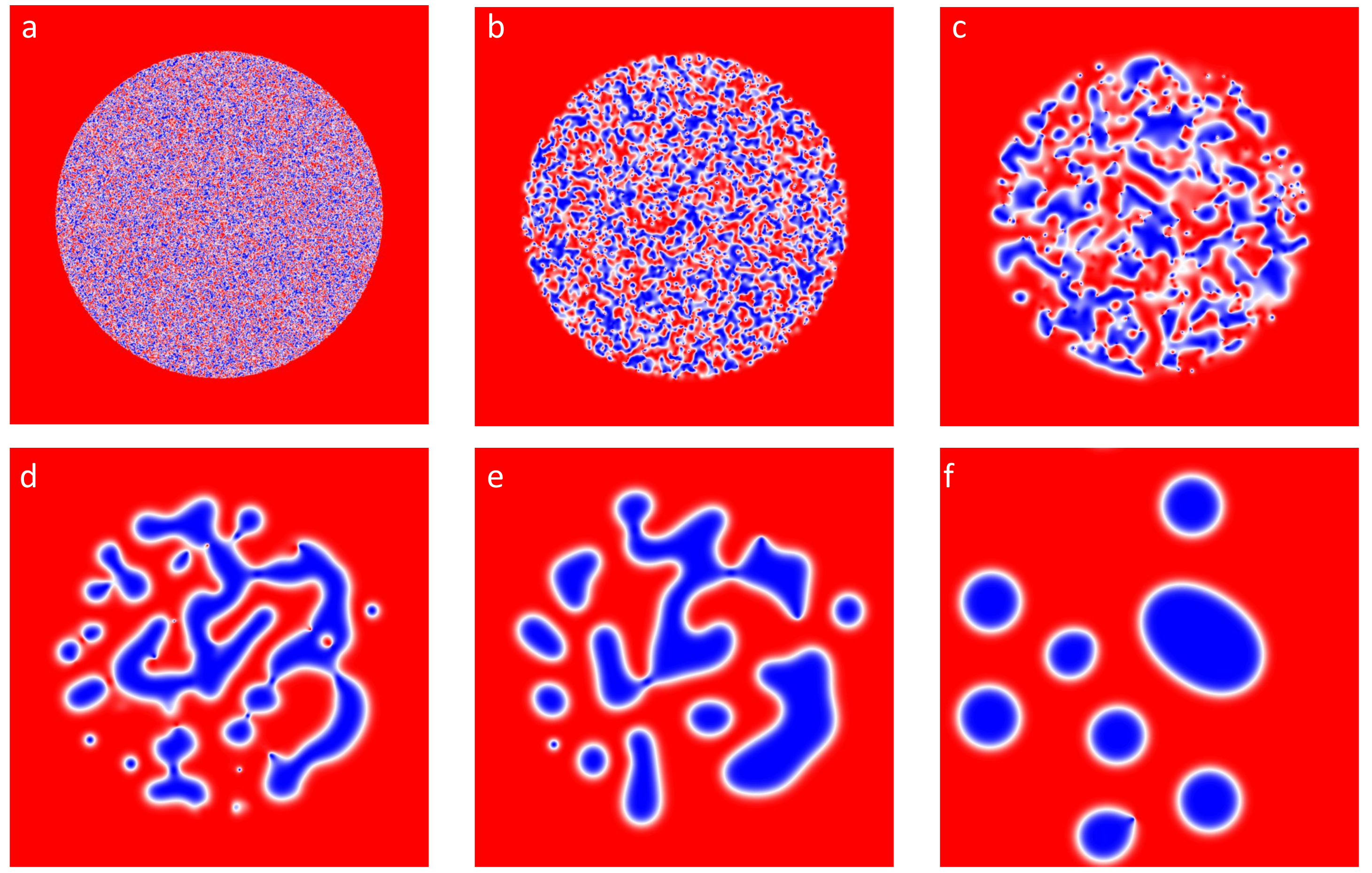}
\centering
\caption{\textbf{Micromagnetic simulations - Laser-induced SAF skyrmion nucleation $\mid$} Micromagnetic simulation of the evolution of magnetization during the minimization of the energy (from \textbf{a} to \textbf{f}), starting from an initial circular demagnetized state mimicking the laser induced demagnetization (\textbf{a}). The simulation is achieved on a $1536\times1536$ nm$^2$ area.}
\label{FIG_laser_simu}
\end{figure}

\section{Additional experiments: observation of chiral domain walls  and skyrmions in a compensated SAF using XMCD-PEEM}
\begin{figure}[h!]
\includegraphics[width=\textwidth]{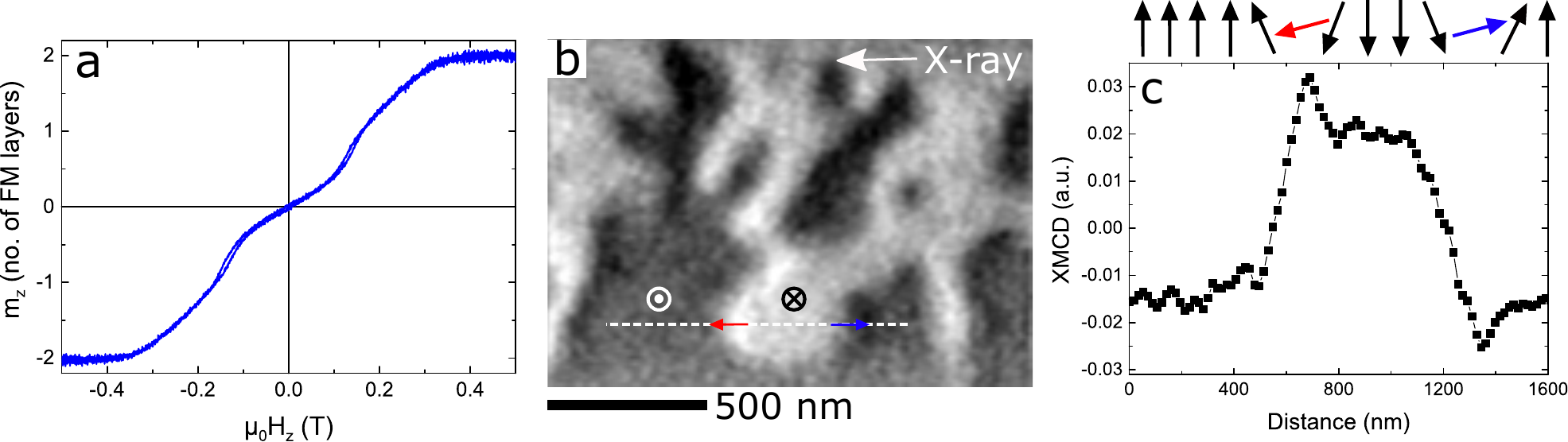}
\centering
\caption{\textbf{Chiral DWs and skyrmion in a compensated SAF $\mid$} \textbf{a.} OOP hysteresis loop measured by VSM on a compensated SAF with two constituent FM layers. \textbf{b.} XMCD-PEEM image acquired at the Co $L_3$ absorption edge and at zero external magnetic field. \textbf{c.} Line-scan of the XMCD contrast along the X-ray beam direction indicated by the white dashed line in \textbf{b}.}
\label{FIG_PEEM}
\end{figure}

In STXM experiments, only the OOP component of the magnetization is accessible. To determine the chirality of the observed spin textures, we performed XMCD-PEEM experiments on a compensated SAF bilayer deposited on a high-resistivity Si wafer. The SAF composition was Ta(3){\slash}Pt(3){\slash}Co(0.2){\slash}Ni$_{80}$Fe$_{20}$(1){\slash}Co(0.2){\slash}Ru(0.85){\slash}Pt(0.5){\slash}Co(0.9){\slash}Ru(0.85){\slash}Ta(1.5) (thickness in nanometers). Fig. \ref{FIG_PEEM}.a shows the OOP hysteresis loop measured by VSM, which exhibits a very similar behavior as the sample described in the previous section. Fig. \ref{FIG_PEEM}.b shows a XMCD-PEEM image acquired at the Co $L_3$ absorption edge and at zero external field. It displays alternate up{\slash}down domains as well as an isolated magnetic skyrmion of diameter $d_{sk}\approx{}160$ nm. In Fig. \ref{FIG_PEEM}.c, we plot the XMCD contrast measured along the X-ray beam at the position marked with the white dashed line in Fig. \ref{FIG_PEEM}.b. Since the contrast is proportional to the projection of the X-ray beam direction on the magnetization, we choose domain walls (DWs) perpendicular to the X-ray beam. The XMCD contrast exhibits the typical maximum and minimum corresponding respectively to a magnetization parallel and anti-parallel to the beam~\cite{boulleRoomtemperatureChiralMagnetic2016}. Hence, the magnetization rotates according to $\uparrow\leftarrow\downarrow\rightarrow\uparrow$, which is the signature of a left-handed N\'eel DW. This strong white{\slash}black contrast is not observed for DWs perpendicular to the beam direction, which further supports that the DWs are of N\'eel type. This left-handed N\'eel character is consistent with the sign of the DMI in both constituent FM layers. Unfortunately, the NiFe layer could not be observed in this experiment since the layer was deeper than the secondary electron escape length. Nevertheless, the AF coupling  promotes DWs of the same chirality in both constituent FM layers so  left-handed N\'eel spin textures are expected in the NiFe layer.

\newpage
\bibliographystyle{apsrev4-2}
\bibliography{MainSuppBib}

\end{document}